\documentclass[letterpaper,twocolumn,10pt]{article}
\usepackage{usenix}

\usepackage{tikz}
\usepackage{hyperref}
\usepackage{booktabs}
\usepackage[ruled,linesnumbered]{algorithm2e}
\usepackage{amsmath,amsfonts,pifont,amsthm}
\usepackage{algorithmic}
\usepackage{tabu}
\usepackage{graphicx}
\usepackage{tabularx,ragged2e}
\usepackage{textcomp}
\usepackage[dvipsnames]{xcolor}
\usepackage[T1]{fontenc}
\usepackage{booktabs} 
\usepackage{multirow, makecell}
\PassOptionsToPackage{table,xcdraw,dvipsnames}{xcolor}
\usepackage{colortbl}
\usepackage{soul}
\usepackage{bm}
\usepackage{hyperref}
\usepackage{subcaption}
\usepackage{latexsym}
\usepackage{mathtools}
\usepackage{enumitem}
\usepackage{xspace}
\usepackage{soul}
\usepackage{fontawesome}
\usepackage{wasysym}
\usepackage{anyfontsize}
\usepackage{xcolor}  
\usepackage{colortbl}  
\usepackage[breakable]{tcolorbox}

\usepackage{soul}
\newcommand{\bluehl}[1]{\sethlcolor{blue!20}\hl{#1}}
\newcommand{\redhl}[1]{\sethlcolor{red!20}\hl{#1}}

\newif\ifshowcomment
\showcommenttrue
\ifshowcomment
    \newcommand{\tabitem}{~~\llap{\textbullet}~~}
    \newcommand{\yujin}[1]{{\color{cyan} {[Yujin: #1]}}}
    
    \newcommand{\update}[1]{{\color{blue} {#1}}}
    \newcommand{\tianneng}[1]{{\color{violet} {[Tianneng: #1]}}}
    \newcommand{\zhun}[1]{\textcolor{violet}{[zhun: #1]}}
    \newcommand{\dawn}[1]{{\color{orange} {[Dawn: #1]}}}
    \newcommand{\kurt}[1]{{\color{teal} {[Kurt: #1]}}}
    
    \newcommand{\andy}[1]{{\color{brown} {[Andy: #1]}}}
\else
    \newcommand{\tabitem}{~~\llap{\textbullet}~~}
    \newcommand{\yujin}[1]{}
    \newcommand{\wenbo}[1]{}
    \newcommand{\update}[1]{}
    \newcommand{\tianneng}[1]{}
    \newcommand{\zhun}[1]{}
    \newcommand{\dawn}[1]{}
    \newcommand{\kurt}[1]{}
    \newcommand{\patrick}[1]{}
    \newcommand{\andy}[1]{}
\fi

\usepackage{authblk}

\begin{document}

\date{}

\title{\Large \bf Frontier AI's Impact on the Cybersecurity Landscape}

\author{
Yujin Potter$^{1*}$, Wenbo Guo$^{2*}$, Zhun Wang$^{1}$, Tianneng Shi$^{1}$, Hongwei Li$^{2}$, Andy Zhang$^{1}$,\\\vspace{-2mm}
Patrick Gage Kelley$^{3}$, Kurt Thomas$^{3}$, and Dawn Song$^{1}$\\\vspace{5mm}
\textsuperscript{1}UC Berkeley \quad
\textsuperscript{2}UC Santa Barbara \quad
\textsuperscript{3}Google
}

\maketitle
\def\thefootnote{*}\footnotetext{Co-first authors}
\def\thefootnote{\arabic{footnote}}
\begin{abstract} 
The impact of frontier AI (i.e.,~\textit{AI agents and foundation models}) in cybersecurity is rapidly increasing. 
In this paper, we comprehensively analyze this trend through multiple aspects: quantitative benchmarks, qualitative literature review, empirical evaluation, and expert survey.
Our analyses consistently show that AI’s capabilities and applications in attacks have exceeded those on the defensive side.
Our empirical evaluation of widely used agent systems on cybersecurity benchmarks highlights that current AI agents struggle with flexible workflow planning and using domain-specific tools for complex security analysis---capabilities particularly critical for defensive applications.
Our expert survey of AI and security researchers and practitioners indicates a prevailing view that AI will continue to benefit attackers over defenders, though the gap is expected to narrow over time.
These results show the urgent need to evaluate and mitigate frontier AI's risks, steering it towards benefiting cyber defenses.
Responding to this need, we provide concrete calls to action regarding: the construction of new cybersecurity benchmarks, the development of AI agents for defense, the design of provably secure AI agents, the improvement of pre-deployment security testing and transparency, and the strengthening of user-oriented education and defenses. 
Our paper summary and blog are available at \url{https://rdi.berkeley.edu/frontier-ai-impact-on-cybersecurity/}.
\end{abstract}

\section{Introduction}
\label{sec:intro}

The security community is increasingly using frontier AI,\footnote{We use frontier AI to represent foundation models like large language models (LLMs) and multi-modal models, as well as AI agents~\cite{durante2024agent}---systems that combine foundation models with symbolic software components.} e.g., using LLMs for fuzzing~\cite{meng2024large} and patching~\cite{zhang2024evaluating}.
Researchers have also developed foundation models for security tasks such as analyzing networking traffic, binary code, and blockchain transactions, achieving better performance than traditional ML in threat detection~\cite{guthula2023netfound,gai2023blockchain,yu2024blockfound,pei2020xda}.
More recently, AI agents from the DARPA AIxCC competition~\cite{darpa} discovered and patched zero-day vulnerabilities~\cite{SQLite3}.
Despite these defensive advances, frontier AI introduces significant risks, particularly its potential for misuse by attackers---ranging from the automation of cyber attacks~\cite{kang2024exploiting,wan2024cyberseceval,fang2024llm,tann2023using} to attacks against AI systems~\cite{wang2023decodingtrust, barrett2023identifying}, e.g, backdoors~\cite{nie2024trojfm}, jailbreaking~\cite{chen2024llm,chen2024rl}, and prompt injection~\cite{liu2023prompt}.

\begin{figure}
    \centering
    \includegraphics[width=1.0\linewidth]{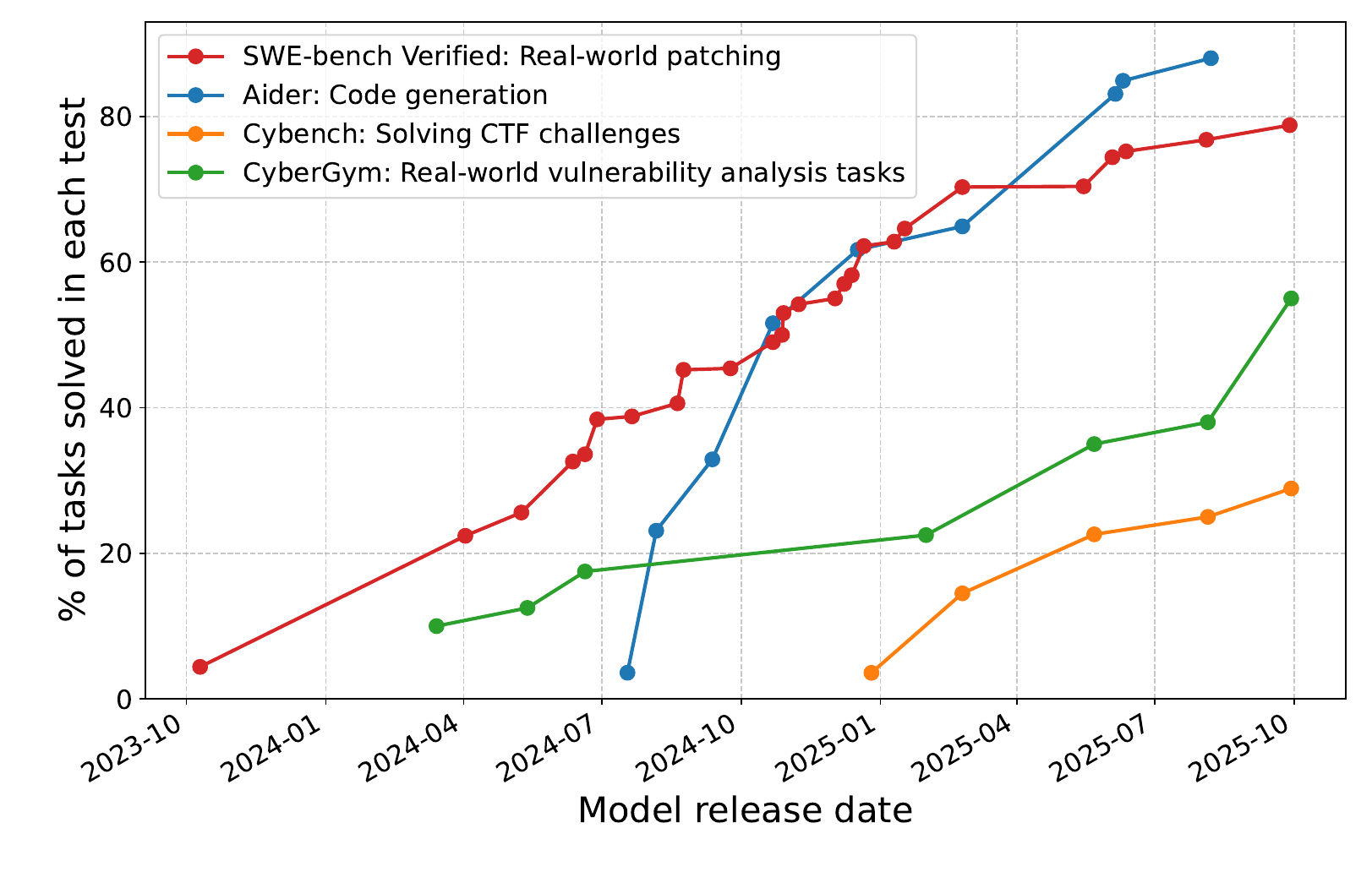}
    \caption{Software engineering performance evolvement of frontier AI over time.}
    \label{fig:trend}
\end{figure}

Given this dual nature, a key question arises: \emph{Will frontier AI provide more benefits to attackers or defenders, and how might it reshape the cybersecurity landscape?}
While several recent papers examine LLMs in cybersecurity~\cite{ferrag2024generative,hassanin2024comprehensive,girhepuje2024survey,yao2024survey,sai2024generative,zhang2025llms,zhou2024large,hou2024large,sheng2025large,schroer2025sok}, these studies do not address these questions: (1) How can we evaluate the current state of frontier AI in attacks and defenses? (2) How might the offense-defense balance change with frontier AI? and (3) What actions need to be taken to ensure AI benefits defense, increasing the security of systems?
Answering these questions is especially critical to prepare for increasing access to frontier AI, amid the unprecedentedly rapid pace of AI evolvement (Figure~\ref{fig:trend})~\cite{aisafetyreport,shah2025approach}.

\begin{figure*}[th]
    \centering
    \includegraphics[width=1.0\linewidth]{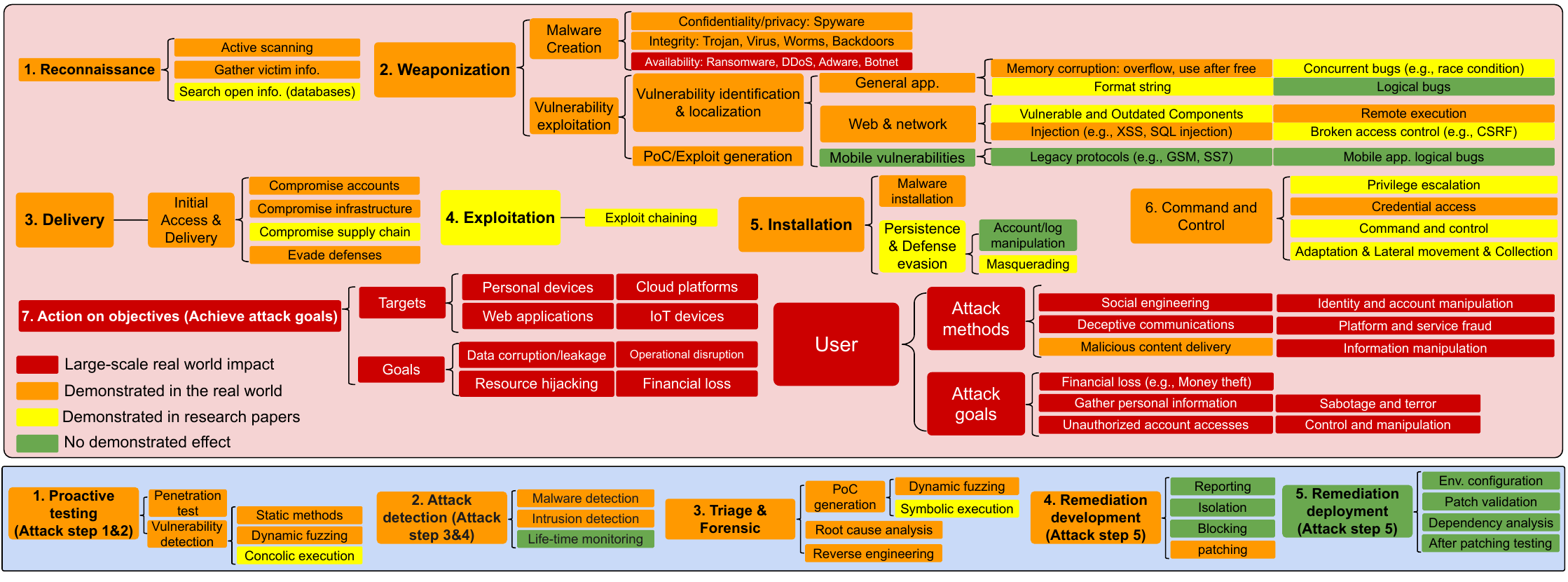}
    \caption{A comprehensive categorization of attacks (red canvas) and defenses (blue canvas) with frontier AI's qualitative impact. AI has a larger impact on final attack goals than on specific stages. The disparity exists because most reports disclose only final impacts rather than attack procedures. For precision, we include only reports with clear specifications for individual attack steps.}
    \label{fig:risk}
\end{figure*}

Our work presents a comprehensive analysis through~\emph{benchmark analysis, literature review, empirical evaluation, and expert survey}.
First, we describe a marginal risk framework~\cite{kapoor2024societal} for analyzing AI's impact in cybersecurity (Section~\ref{sec:methodology}). 
We then apply this taxonomy to assess frontier AI’s current impacts on both attacks and defenses quantitatively and qualitatively (Sections~\ref{sec:benchmark} and \ref{sec:qualitative}). Our analysis indicates that \emph{AI has been utilized more in attacks---particularly in human-targeted attacks---than in defense, enabling greater automation and facilitation to attackers.}

We also empirically evaluate the state-of-the-art (SOTA) coding agent OpenHands~\cite{wang2024openhands} on the popular attack and defense benchmarks (Section~\ref{sec:eavl}).
The results support our earlier conclusion and further uncover~\emph{the limitations of current agents in architecture and workflow planning}.
Specifically, the agent is not equipped with domain-specific tools, especially those beneficial for defenses, which prevents it from conducting security analyses on large codebases.
As a result, the agent performs poorly in difficult tasks, such as exploit development, proof-of-concept (PoC) generation, and root cause analysis (if not in stack traces).

Through the expert survey, we also find that experts believe that frontier AI will benefit attackers more, even though the gap is expected to narrow over time; within the next two years, the balance is expected to favor attackers by $31.7\%,$ within five years by $12.8\%,$ and within ten years by just $0.46\%$ (Section~\ref{sec:survey}). Our position aligns with this view, which we support through analysis of the equivalence class problem and the fundamental asymmetry between attackers and defenders (Section~\ref{sec:short}).

Finally, we offer actionable recommendations to guide the community as follows (Section~\ref{sec:suggestions}):
\begin{itemize}
    \item We recommend building security benchmarks with real-world systems and executable environments for agents.
    \item We recommend improving traditional defenses with AI-based rule extraction and automation, as well as constructing AI agents with security-specific tools that automatically determine their workflows.
    \item We recommend AI developers conduct comprehensive pre-deployment security testing and enforce greater transparency in AI development.
    \item We call for better education about AI-based attacks and security best practices, as well as developing human-oriented defenses.
\end{itemize}

\noindent In Appendix~\ref{app:open}, we list a set of open research questions about frontier AI and security. 

\section{Methodology}
\label{sec:methodology}

\noindent\textbf{Marginal risk framework and scope.}
Based on the concept of marginal risk~\cite{kapoor2024societal}---the new/unique risks beyond the existing risks---we define the \textit{marginal risks of frontier AI in cybersecurity} as novel cyber threats either introduced or amplified by frontier AI through intentional misuse beyond traditional attack methods.
For attacks against systems, we categorize attacks into seven steps based on the MITRE framework~\cite{mitre_attack_2024} and the cyber kill chain framework~\cite{kill_chain_2024} (Figure~\ref{fig:risk}); within each step, we define subcategories by activity and application.
Similarly, we classify attacks against humans based on their goals and methods, as well as defenses for systems.
We acknowledge that covering all risks and attack vectors is challenging; therefore, we limit our scope to cyberattacks against systems and humans (e.g, we do not discuss cryptographic attack vectors).
With the taxonomy, we study frontier AI's impact as follows:

\smallskip
\noindent\textbf{Quantitative benchmarking.} 
Because the evaluation of frontier AI’s cybersecurity capabilities is still at an early stage, available benchmarks remain limited. 
To collect benchmarks, we begin with several early and representative cybersecurity benchmarks (e.g., Cybench~\cite{zhang2024cybench}), selected based on our domain expertise. 
We expand this set by adding other benchmarks cited within those works. 
We compile $34$ cybersecurity benchmarks and categorize them based on our taxonomy. 
We also identify the limitations of current benchmarks and propose calls-to-action to improve benchmarking in Section~\ref{sec:suggestions}.

\smallskip
\noindent\textbf{Literature review.}
We conduct a literature search using keywords related to frontier AI and cybersecurity in Google Scholar.
The scope includes peer-reviewed papers, preprints, industry white papers, reports of real-world attack instances, technical blogs, and discussions from social and community platforms (e.g., X/Twitter) published between 2021 and Aug. 2025. This search produces an initial list of over 500 papers. In selecting articles, we prioritize conference publications, preprints from leading AI labs and major companies, and works with high citation impact.
Five researchers participated in this process, resulting in a final list of $183$ papers. 
Through a qualitative analysis, we then map each attack and defense step to a four-tier impact:
(1) \textit{no demonstrated effect}, where we find no AI-related impact for this step;
(2) \textit{demonstration in research papers}, where we find only research papers but no real-world reports;
(3) \textit{demonstration in real-world attacks}, where we find research papers and a few early-stage real-world demonstrations; 
(4) \textit{large-scale real-world deployments}, where AI has been widely deployed or had a noticeable real-world impact.

\smallskip
\noindent\textbf{Empirical Evaluation.} To examine the limited defensive capabilities of current agents relative to their offensive capabilities, we evaluate the offensive and defensive capabilities of the SOTA agent, OpenHands~\cite{wang2024openhands}, using several cybersecurity benchmarks. For offensive capability, we employ AutoPenBench~\cite{gioacchini2024autopenbench}, which includes end-to-end attack scenarios. For defensive capability, we use CyberGym~\cite{wang2025cybergym} for PoC generation and PatchAgent~\cite{yu2025patchagent} for patch generation. Detailed experimental setups and designs are described in Section~\ref{sec:eavl}.

\smallskip
\noindent\textbf{Expert survey.}
To assess how the AI/security community perceives the current and future impacts of frontier AI on cybersecurity, we conducted an expert survey targeting AI/security researchers and practitioners. The survey consists of questions about the current status quo of offensive and defensive use of frontier AI, predictions regarding timelines for automating cybersecurity tasks, and expectations about how frontier AI may shift the balance between attackers and defenders. 
We used snowball sampling to share the survey with experts; we emailed colleagues working at the intersection of AI and cybersecurity in academia and industry and asked them to circulate the survey within their professional networks. 
In total, $129$ experts accessed the survey, of whom $34$ completed it. Some questions were randomly assigned to respondents as part of the dynamic survey design, resulting in varying sample sizes across questions, with a maximum of $n=46.$ Survey questions, per-question sample sizes, and demographic characteristics (for respondents who provided them) are presented in Appendix~\ref{app:survey}.
We analyze and report experts’ forecasts of frontier AI’s future impacts and their views on the evolving offense-defense balance, including whether--and how--the perceived advantage for attackers may narrow over time.

\section{Quantitative Analysis: Benchmarks}
\label{sec:benchmark}

\begin{table*}[!ht]
\caption{Overview of existing benchmarks. Recon. - Reconnaissance, Mal. C - Malware creation, Vuln. E - Vulnerability exploitation, Ini. A \& Del. - Initial access \& delivery, Ex. \& Inst. - Exploitation \& installation, Per. \& Ev. - Persistence \& evasion, C2. \& Act. - Command and control \& action on objectives, Pen. Test - Penetration test, Vul. D - Vulnerability detection, Att. D - Attack detection, PoC \& RT - PoC \& Root cause, Rem. Dev. - Remediation development, Rem. Dep. - Remediation deployment. *Since Q\&A benchmarks focus mainly on evaluating models' cybersecurity knowledge, we excluded them from the table.}
\centering
\resizebox{\textwidth}{!}{
\begin{tabular}{r|r|r|c|c}
\Xhline{1.0pt}
\multicolumn{2}{r|}{\textbf{Attack steps}} & \textbf{Quantity} & \textbf{Notable examples} & \textbf{Key limitations} \\ \hline
1 & Recon. & 1 & AutoPenBench~\cite{gioacchini2024autopenbench} & Limited method and scenario coverage; limited agent scaffolds \\ \cline{1-5}
\multirow{2}{*}{\begin{tabular}[r]{@{}r@{}}2 \end{tabular}} & Mal. C & 2 & RedCode~\cite{guo2024redcode}, CySecBench~\cite{wahreus2025cysecbench} & Limited metric for malware creation, \\ \cline{2-4}
 & Vul. E & 7 & CyberSecEval~\cite{wan2024cyberseceval}, BaxBench~\cite{vero2025baxbench},  BountyBench~\cite{zhang2025bountybench}, CyberGym~\cite{wang2025cybergym} &  limited program language coverage \\ \cline{1-5}
\multirow{3}{*}{\begin{tabular}[r]{@{}r@{}}3--5 \end{tabular}} & {Ini. A \& Del. } & 2  &  AutoPenBench~\cite{gioacchini2024autopenbench}, SecCodePLT~\cite{yang2024seccodeplt} & Limited attack category and type coverage, \\ \cline{2-4} 
& {Ex. \& Inst.} & 3 & CyBench~\cite{zhang2024cybench}, SecCodePLT~\cite{yang2024seccodeplt} & limited benchmarks on real-world complex systems, \\ \cline{2-4} 
& {Per. \& Ev. } & 0 &  None & limited dynamic benchmarks with agent scaffold \\ \cline{1-4} 
6--7 & {C2. \& Act.} & 1 & SecCodePLT~\cite{yang2024seccodeplt} &  \\ 
\Xhline{1.0pt}
\multicolumn{2}{r|}{\textbf{Defense steps}} & \textbf{Quantity} & \textbf{Notable examples} & \textbf{Key limitations} \\ \hline
\multirow{2}{*}{\begin{tabular}[r]{@{}r@{}}1 \end{tabular}} & Pen. test & 1 & AutoPenBench~\cite{gioacchini2024autopenbench} & Limited attacks, target systems, metrics, \\ \cline{2-4}
& Vul. D & 12 &PrimeVul~\cite{ding2024vulnerability}, SVEN~\cite{he2023large} FuzzBench~\cite{metzman2021fuzzbench}, OSS-Fuzz~\cite{serebryany2017oss} & limited complexity and diversity, low-quality labels\\ \hline
2 & Att. D & 4 & IDS2018~\cite{sharafaldin2018toward}, Drebin~\cite{arp2014drebin}, BODMAS~\cite{yang2021bodmas} & Low-quality data and labels, lack OOD/hidden tests \\ \hline
\multirow{1}{*}{\begin{tabular}[r]{@{}r@{}} 3 \end{tabular}} & {PoC \& RT} & 2 & CRUXEval~\cite{gu2024cruxeval}, CyberGym~\cite{wang2025cybergym} & No benchmark for root cause   \\ \hline
4 & Rem. Dev. & 5 & SWE-bench~\cite{jimenez2023swe}, SecCodePLT~\cite{yang2024seccodeplt}, BountyBench~\cite{zhang2025bountybench}  & Limited benchmarks for vuln. in different languages\\ \hline
5 & Rem. Dep. & 0 & None & N/A \\ \Xhline{1.0pt}
\end{tabular}
}
\label{tab:bench}
\end{table*}

\subsection{Offensive Benchmarks}
\label{subsec:attack_benchmark}

\emph{As shown in Table~\ref{tab:bench}, except for vulnerability exploitation in step 2, existing benchmarks are limited in risk coverage, metric design, and dynamic evolution.}
Table~\ref{tab:llm_attack_bench} in Appendix details SOTA AI's performances on these attack benchmarks.

\smallskip
\noindent\textbf{Attack Step 1: Reconnaissance.}
Early benchmarks focus on question answering~\cite{tihanyi2024cybermetric,liu2024cyberbench,wan2024cyberseceval,kouremetis2025occult}. 
AutoPenBench~\cite{gioacchini2024autopenbench} enables attack generation and execution but covers limited methods and scenarios.
Results show that simple AI agents can handle reconnaissance like discovering potential victims on remote networks or conducting surveillance.
However, agents cannot identify target-specific vulnerable services due to limited service information from standard tools (e.g., Nmap) and a lack of deeper interactions with target systems. 

\smallskip
\noindent\textbf{Attack Step 2: Weaponization.}
For malware creation, RedCode~\cite{guo2024redcode} tests LLMs' capabilities in generating malware in Python and uses VirusTotal as the judge.
CySecBench~\cite{wahreus2025cysecbench} tests the safety alignment of LLMs and shows that Claude 3.5 Sonnet is more resilient than GPT-4o and Gemini-1.5. 
Recent works design benchmarks for insecure code generation (e.g., CyberSecEval~\cite{wan2024cyberseceval}, SecCodePLT~\cite{yang2024seccodeplt}, BaxBench~\cite{vero2025baxbench}, SecRepoBench~\cite{dilgren2025secrepobench}).
They craft specific coding tasks that may encounter known CWEs and evaluate whether LLMs generate vulnerable code when finishing these tasks, showing that both SOTA commercial and open-source models can generate vulnerable code, with stronger models being more vulnerable due to their enhanced task understanding and execution.
BountyBench~\cite{zhang2025bountybench} and CVE-Bench~\cite{zhu2025cvebench} evaluate agent performance on vulnerability exploitation, showing that agents using Claude 3.7 Sonnet successfully exploit $67.5\%$ of tasks on BountyBench, while agents using GPT-4o achieve a $13\%$ success rate on CVE-Bench.
CyberGym~\cite{wang2025cybergym} evaluates agents' capabilities in PoC generation on real-world software projects, identifying $15$ zero-day vulnerabilities.
All existing benchmarks focus on C/C++ with limited coverage of other languages. 


\smallskip
\noindent\textbf{Attack Steps 3$\sim$7.}
Benchmarks for these later attack steps remain severely limited in coverage and quality.
Similar to reconnaissance, most existing benchmarks are about Q\&A rather than generating actual attacks~\cite{tihanyi2024cybermetric,liu2024cyberbench,wan2024cyberseceval,kouremetis2025occult}.
SecCodePLT~\cite{yang2024seccodeplt} and AutoPenBench~\cite{gioacchini2024autopenbench} include partial attack steps and use dynamic testing as the judge.
SecCodePLT~\cite{yang2024seccodeplt} provides a network system with pre-specified attack paths and targets, enabling end-to-end attacks, including reconnaissance, initial control, C2, and collection.
Their results show that Claude 3.5 Sonnet's safety alignment rejects most attack generation queries, while GPT-4o generates end-to-end attacks with low success rates.
CTF benchmarks~\cite{shao2024empirical,zhang2024cybench} include pwn challenges testing privilege escalation, showing that commercial LLMs slightly outperform open-source ones.

\subsection{Defensive Benchmarks}
\label{subsec:defense_benchmark}

\emph{Most benchmarks focus on vulnerability detection and attack detection, while other defense steps lack benchmarks.}
Table~\ref{tab:llm_defense_bench} in Appendix details AI's performance on these benchmarks.

\smallskip
\noindent\textbf{Defense Step 1: Proactive testing.}
Benchmarking penetration testing is related to attack benchmarking, which remains at an early stage with AutoPenBench~\cite{gioacchini2024autopenbench}.
Existing vulnerability detection benchmarks include static benchmarks with labeled functions~\cite{pearce2022asleep,ding2024vulnerability,ullah2024llms,chauvin2024eyeballvul,tony2023llmseceval,vulhub,Hackthebox,owasp} and dynamic benchmarks with fuzzing instrumentation and execution environment~\cite{serebryany2017oss,metzman2021fuzzbench}.
For static benchmarks, most existing benchmarks remain at the individual function level, e.g.,  PrimeVul~\cite{ding2024vulnerability} and SVEN~\cite{he2023large}.
These datasets have low-quality labels due to missing context (functions vulnerable only under conditions specified by other functions in the projection).
SecCodePLT~\cite{yang2024seccodeplt} includes a project-level vulnerability detection dataset for C/C++ and Java, showing the limited performance of existing models and agents in these languages.
For dynamic benchmarks, OSS-Fuzz~\cite{serebryany2017oss} is the largest continuous fuzzing platform for critical open-source projects.
FuzzBench~\cite{metzman2021fuzzbench} is a smaller but less noisy benchmark widely used in research papers to test novel fuzzing methods, including AI-facilitated fuzzers~\cite{shi2024harnessing,zhang2025g2fuzz}. 

\smallskip
\noindent\textbf{Defense Step 2: Attack detection.}
Many benchmarks exist for network intrusion detection and malware detection~\cite{sharafaldin2018toward,arp2014drebin,yang2021bodmas}.
Recent research~\cite{arp2022and} highlights these benchmarks' limitations in data quality (duplicated data, shortcuts) and label accuracy.
Transformers show very high accuracy on these benchmarks~\cite{guthula2023netfound,lin2022bert}.
Besides attack detections, CTIBench~\cite{alam2024ctibench} designs tasks for LLMs to pinpoint the specific CWEs and attack techniques for given attack instances.

\noindent\textbf{Defense Step 3: Triage \& forensic.}
CRUXEval~\cite{gu2024cruxeval} benchmarks PoC generation for vulnerable C/C++ codes with overflow bugs, showing that GPT-4o achieves 75\% pass@1 success rate. 
CyberGym~\cite{wang2025cybergym} offers more challenging tasks for agents, including PoC generation on real-world codebases with vulnerability descriptions, showing that OpenHands with Claude 4 Sonnet achieves a 17.9\% success rate.

\smallskip
\noindent\textbf{Defense Steps 4\&5: Remediation development and deployment.}
\emph{Given that AI is not yet widely involved in these steps, there are not many benchmarks.} 
The most popular benchmark is SWE-bench~\cite{jimenez2023swe}, which extracts issues from multiple GitHub Python projects. 
SOTA multi-agent systems with GPT-5 and Claude 4 Sonnet can resolve more than $70\%$ issues in the SWE-bench-verified benchmark. 
Recent works extend the SWE-bench to multi-modal settings~\cite{yang2024swe} and the Java language~\cite{zan2024swe}. 
SecCodePLT~\cite{yang2024seccodeplt} (Python, C/C++, Java) and BountyBench~\cite{zhang2025bountybench} (C, Python, JavaScript/TypeScript) include benchmarks of security patches.
Both benchmarks use unit tests to test functionality. 
SecCodePLT tests security via fuzzing through PoC, and BountyBench tests whether the patch prevents human-written exploits.
Top agents (e.g., OpenAI Codex CLI o4-mini) can achieve high results on BountyBench (>90\%), but remain ineffective on SecCodePLT (<30\%).

\begin{table*}[!t]
    \centering
    \caption{Key summaries of the qualitative literature review. The colors in the steps correspond to the same as in Figure~\ref{fig:risk}. Literature supporting research impact and real-world impact is highlighted in blue and red, respectively.}
    \resizebox{\textwidth}{!}{
    \begin{tabular}{>{\centering\arraybackslash}p{0.1\textwidth}|>{\centering\arraybackslash}p{0.2\textwidth}|>{\centering\arraybackslash}p{0.24\textwidth}|p{0.5\textwidth}}
        \Xhline{1.0pt}
       \textbf{Attack steps} & \textbf{Description} & \multicolumn{1}{c|}{\textbf{Papers}} & \multicolumn{1}{c}{\textbf{Impact Summary}} \\
       \hline\hline
        \cellcolor{orange!80} 1: Reconnaissance & Scan the env. or harvest info. to identify targets & \bluehl{\mbox{\cite{sun2024gptscan,deng2024pentestgpt,nvidia_cve_analysis}}}\redhl{\mbox{\cite{Microsoft}}} 
        & LLMs in reconnaissance include active scanning, victim information gathering, and open-source database search. \\
        \hline
        \cellcolor{orange!80}Step 2: Weaponization & Create malware and exploits and encapsulate them as weapons &\multirow{1}{*}{\makecell{\bluehl{\mbox{\cite{pa2023attacker,sheng2025large,lu2024grace,li2024llm,sakaoglu2023kartal,ding2024vulnerability}}}\\\bluehl{\mbox{\cite{liumake,wan2024cyberseceval,yang2023exploitgen,wang2025cybergym,fang2024llm,zhang2025bountybench}}}\\\bluehl{\mbox{\cite{fang2024teams,patil2024leveraging,tann2023using,zhang2024cybench,shao2024empirical}}}\\\redhl{\mbox{\cite{ddos,a10ddos2024,Microsoft,SQLite3,imperva_news,google_bigsleep}}}}}
        & Research papers show LLMs can generate functional malware with high evasion rates and AI agents aid in vulnerability identification and exploitation, although large-scale real-world attacks remain limited to simpler attacks like DDoS and SQL injection.\\
        \hline
        \cellcolor{orange!80} Step 3: Delivery & Transmit attacks through e-mail, USB, web &\bluehl{\mbox{\cite{hitaj2019passgan,wang2023password,securityhero2023,plesner2024breaking,deng2024oedipusllmenchancedreasoningcaptcha,debicha2023advbot}}} \redhl{\mbox{\cite{gmail_account_news,openai_report_disrupting,imperva_news,Microsoft}}} & Initial access are demonstrated in the real world, with AI assisting in password guessing and bypassing defenses such as 2FA. \\
        \hline
        \cellcolor{yellow!80} Step 4: Exploitation & Exploit vulnerabilities in the victim & \bluehl{\mbox{\cite{de2024chainreactor}}} & AI can write simple exploits but cannot craft exploit chains. \\
        \hline
        \cellcolor{orange!80} Step 5: Installation & Install the malware on the victim’s system& \bluehl{\mbox{\cite{beckerich2023ratgpt}}}\redhl{\mbox{\cite{google_adv_misuse,deeplocker}}} & AI is used for malware installation but not persistence in the real world. \\
        \hline
        \cellcolor{orange!80} Step 6: Command \& control & Establish a remote command channel & \multirow{1}{*}{\makecell{\bluehl{\mbox{\cite{happe2024llmshackersautonomouslinux,de2024chainreactor,tulla2025alfa,kujanpaa2021automating,happe2023evaluating,deeplocker,anderson2016deepdga}}}\\\bluehl{\mbox{\cite{beckerich2023ratgpt,chung2019availability}}}\redhl{\mbox{\cite{obviam-password-cracking,securityhero-password-cracking,cepa-ai-cyberattacks}}}}} & Privilege escalation through exploit chain generation and command\&control through automated domain generation are demonstrated in research. Credential access is used in the real world.\\
        \hline
        \cellcolor{BrickRed!100} Step 7: Action on objectives & Achieve their ultimate objectives (e.g.\ data theft, sabotage) through the C\&C channel & \multirow{1}{*}{\makecell{\bluehl{\mbox{\cite{ yang2024seccodeplt,xu2024autoattacker,wang2024sands,usman2024generative,singer2025feasibility}}}\\\redhl{\mbox{\cite{ddos, hp_news, a10ddos2024, imperva_news, gmail_account_news, openai_report_disrupting, crowdstrike}}}\\\redhl{\mbox{\cite{ zero,gartner_survey,bugcrowd_inside,ai-retail,ai-retail2}}}}} &
        Real-world AI-enhanced attacks increase across various systems (Web, mobile, cloud), with malicious purposes including malware deployment, business logic abuse, and credential theft, causing significant financial losses and data breaches.\\
        \hline
        \cellcolor{BrickRed!100} Attacks against humans & Exploit people often through deception, coercion, or manipulation &\multirow{1}{*}{\hspace{-2mm}\makecell{\bluehl{\mbox{\cite{phishing,eze2024analysis,bethany2024large,heiding2024evaluating,lawenforcggment,spearphishing,anthropic_persuasiveness,goldstein2024persuasive,costello2024durably}}}\\\bluehl{\mbox{\cite{potter2024hidden}}}\redhl{\mbox{\cite{hongkong,darktrace,voice_phishing,yao2024survey,fraudgpt,blauth2022artificial}}}\\\redhl{\mbox{\cite{theft,openai_report_disrupting_25feb,fakereview,sullivan2024ai,cyberbullying,psy_control,suicide}}}}} & Frontier AI significantly escalates attacks against humans (e.g., studies showing increases in social engineering and voice phishing since ChatGPT's adoption).\\
             \Xhline{1.0pt}
       \textbf{Defense steps} & \textbf{Description}  & \textbf{Papers} & \multicolumn{1}{c}{\textbf{Impact Summary}} \\
       \hline\hline
       \cellcolor{orange!80} Step 1: Proactive testing & Probe systems before an attack to uncover vulnerabilities &\multirow{1}{*}{\hspace{-2mm}\makecell{\bluehl{\mbox{\cite{kong2025vulnbot,happe2023getting,xu2024autoattacker,temara1maximizing,constantin2023llms,hilario2024generative,isozaki2024towards}}}\\\bluehl{\mbox{\cite{deng2024pentestgpt,zhou2019devign,xu2017neural,ding2024vulnerability,akuthota2023vulnerability,nana2024deep,sakaoglu2023kartal}}}\\\bluehl{\mbox{\cite{lu2024grace,li2024llm,shestov2024finetuning,purba2023software,du2024vul,sun2024llm4vuln,lekssays2025llmxCPG}}}\\\bluehl{\mbox{\cite{sheng2025large,yang2023kernelgpt,oliinyk2024fuzzing,meng2024large,deng2023large,zhang2024llamafuzz,zhang2025g2fuzz}}}\\\bluehl{\mbox{\cite{huang2024large,li2025iris,shi2024bandfuzz,meng2024largehybrid}}}\redhl{\mbox{\cite{zero,repoaudit2025}}}}} &Research explores LLMs for proactive testing, including automated penetration testing and vulnerability detection and AI-enhanced fuzzing. Real-world demonstrations exist but lack evidence of large-scale adoption.\vspace{4mm}\\
       \hline
       \cellcolor{orange!80} Step 2: Attack detection & Identify malicious behavior as it happens and raise real-time alerts & \multirow{1}{*}{\hspace{-2mm}\makecell{\bluehl{\mbox{\cite{wang2017adversary,wu2023grim,caviglione2020tight,vinayakumar2019robust,kwon2019survey,liu2025analyzing,lin2022bert}}}\\\bluehl{\mbox{\cite{al2024exploring,guthula2023netfound,yu2024blockfound,gai2023blockchain,stein2024towards,zahan2024shifting}}}\redhl{\mbox{\cite{microsoft_ire}}}}}& LLMs remove manual feature engineering, reduce reliance on labeled datasets, and generalize better to out-of-distribution cases, with real-world use in malware and intrusion detection.\\
       \hline
       \cellcolor{orange!80} Step 3: Triage \& forensic & Reproduce the vulnerabilities and identify root causes & \multirow{1}{*}{\hspace{-2mm}\makecell{\bluehl{\mbox{\cite{wang2024python,jingxuan2021learch,wang2025cybergym,nitin2025faultline,guo2019deepvsa,mu2019renn,cui2018rept}}}\\\bluehl{\mbox{\cite{roy2024exploring,yang2024large,rafi2024enhancing,shin2015recognizing,chua2017neural,pei2020xda,xie2024resym}}}\\\bluehl{\mbox{\cite{jiang2025beyond,xu2025unleashing,hu2024degpt}}}\redhl{\mbox{\cite{charlotte}}}}} & For PoC generation, recent works mainly explore LLM-assisted fuzzing. AI agents are developed for root cause analysis. Foundation models also show superior performance in binary analysis.\\
       \hline
       \cellcolor{orange!80} Step 4: Rem. dev. & Create patches or other countermeasures & \multirow{1}{*}{\hspace{-4mm}\makecell{\bluehl{\mbox{\cite{li2025sok,yu2024security,kulsum2024case,ahmed2023better,meng2024empirical,sobania2023analysis,chen2024large,pearce2023examining}}}\\\bluehl{\mbox{\cite{zhang2024evaluating,bouzenia2024repairagent,lin2024vulnerabilities,nong2025appatch,yu2025patchagent,li2025patchpilot}}}\redhl{\mbox{\cite{SQLite3}}}}}& Research demonstrates AI can automatically generate security vulnerability patches, but real-world applications remain limited \\
       \hline
       \cellcolor{ForestGreen!80} Step 5: Rem. dep. & Deploy those fixes in production & NA & No specific work exists on AI-assisted remediation deployment.\\
        \hline
        \cellcolor{orange!80}Defense for humans & Protect people from manipulation or exploitation in attacks& \multirow{1}{*}{\makecell{\bluehl{\mbox{\cite{ferrara2023social,zanke2023ai,dhieb2020secure,cao2024phishagent,rana2022deepfake,al2023gpt,anirudh2023multilingual}}}\\\bluehl{\mbox{\cite{dhiman2024gbert,liu2024preventing,zhang2024remark, pan2024markllm, liu2024survey,zhao2024sok, gunn2024undetectable}}}\\\redhl{\mbox{\cite{nvdia,granny}}}}} & Even though frontier AI shows promise for enhancing defenses against human-targeted attacks, defensive techniques struggle to keep pace with sophisticated attacks.\\
        \Xhline{1.0pt}
    \end{tabular}
    }
    \label{tab:summary}
\end{table*}

\section{Qualitative Analysis: Literature Review}
\label{sec:qualitative}

Table~\ref{tab:summary} summarizes the qualitative impact of frontier AI in attacks and defenses along with the categorized papers.

\subsection{Attacks Targeting Systems}
\label{sec:attacks}

AI’s influence on most steps remains at research, though several demonstrate large-scale real-world effects (Figure~\ref{fig:risk}). 

\smallskip
\noindent\textbf{Attack Step 1: Reconnaissance.}
\emph{Real-world attacks increasingly use LLMs and AI agents for active scanning (e.g., satellite capability understanding by Forest Blizzard) and information gathering about victims (e.g., by Salmon Typhoon)}~\cite{Microsoft}. 
Recent research explores frontier AI in multiple reconnaissance methods: active scanning~\cite{sun2024gptscan}, victim information gathering~\cite{deng2024pentestgpt}, and open-source database search~\cite{nvidia_cve_analysis}.

\noindent\textbf{Attack Step 2: Weaponization.}
\emph{Malware generation mainly stays at the research level, where real-world incidents mainly report DDoS attacks~\cite{ddos,a10ddos2024}.}
Specifically, research shows that LLMs can generate Windows PE malware with high evasion rates against VirusTotal~\cite{pa2023attacker} ($>$70\% for Ransomware, Worm, and DDoS); LLM alignments can be evaded~\cite{deng2023jailbreaker,luo2024jailbreakv}; LLMs are also fine-tuned to generate attacks~\cite{setak2024fine}.
For vulnerability identification, LLMs are used in various network and software systems~\cite{sheng2025large,lu2024grace,li2024llm,sakaoglu2023kartal,ding2024vulnerability,liumake}.
Real-world attacks also begin employing LLMs for vulnerability detection, including MSDT vulnerability (CVE-2022-30190)~\cite{Microsoft} and SQLite overflow bugs~\cite{SQLite3}.
Moreover, research shows LLMs can generate PoCs for memory corruption bugs in C/C++~\cite{wan2024cyberseceval,yang2023exploitgen,wang2025cybergym}.
AI agents can exploit real-world N-day~\cite{fang2024llm,zhang2025bountybench} and zero-day~\cite{fang2024teams,patil2024leveraging,zhu2025cvebench} vulnerabilities (e.g., SQL/XSS injection, concurrency attacks, remote code execution) and solve CTF challenges~\cite{tann2023using,zhang2024cybench,shao2024empirical}.
However, \emph{the study on exploitation remains mainly at a research level, with only a few public real-world reports~\cite{imperva_news,Microsoft,google_bigsleep}}.

\noindent\textbf{Attack Steps 3-5: Delivery \& exploitation \& installation.}
\emph{Initial access and installation have been demonstrated in the real world~\cite{openai_report_disrupting,imperva_news}, while persistence remains at a research level~\cite{deeplocker}}.
Initial access is achieved through compromised accounts or infrastructure~\cite{gmail_account_news,openai_report_disrupting,imperva_news}, with AI assisting in password guessing~\cite{hitaj2019passgan,wang2023password,securityhero2023} and bypassing defenses, including intrusion detection, CAPTCHA, and two-factor authentication~\cite{plesner2024breaking,deng2024oedipusllmenchancedreasoningcaptcha,Microsoft,debicha2023advbot}.
LLMs and AI agents can also assist malware configuration and installation~\cite{google_adv_misuse}.
Research works employ frontier AI to maintain persistence across system restarts and credential changes, e.g., AI helps masquerading by disguising malicious activities as benign~\cite{deeplocker,beckerich2023ratgpt}.

\noindent\textbf{Attack Step 6: Commend and control (C\&C).}
Research has demonstrated frontier AI can facilitate privilege escalation~\cite{happe2024llmshackersautonomouslinux,de2024chainreactor,kujanpaa2021automating,happe2023evaluating,tulla2025alfa}.
For instance, ChainReactor~\cite{de2024chainreactor} uses LLM planning to generate exploit chains for privilege escalation.
For stable and stealthy C\&C channels, attackers can employ LLMs for automated domain generation, creating resilient remote management of compromised systems~\cite{deeplocker,anderson2016deepdga,beckerich2023ratgpt}.
Recent research shows LLMs can assist with adaptation~\cite{chung2019availability}, allowing self-learning malware to create environment-tailored attacks.
Only a few real-world reports exist for credential access in this step~\cite{obviam-password-cracking,securityhero-password-cracking,cepa-ai-cyberattacks}.

\noindent\textbf{Attack Step 7: Action on objectives.}
\emph{AI-enhanced cyberattacks have proliferated across web, IoT, mobile, and cloud platforms, causing real-world financial losses, privacy leakage, data manipulation, and resource hijacking}~\cite{ddos, hp_news, a10ddos2024, imperva_news, gmail_account_news, openai_report_disrupting, crowdstrike, zero, gartner_survey, bugcrowd_inside,yang2024seccodeplt,singer2025feasibility}.
A recent attack deploys AsyncRAT malware using LLMs to leak sensitive information~\cite{hp_news}.
Frontier AI amplifies attacks against web applications and IoT devices in e-commerce, including business logic abuse, API violations, bad bot attacks, and targeted DDoS on retail sites~\cite{ai-retail,ai-retail2}.
These attacks disrupt operations, steal data, and exploit digital engagement for financial gain~\cite{imperva_news}.
CrowdStrike reports a 110\% rise in cloud-aware cyberattacks and 75\% increase in intrusions year-over-year with AI enhancement~\cite{crowdstrike}.
Early research explores end-to-end attacks using LLMs or AI agents~\cite{xu2024autoattacker,wang2024sands,usman2024generative,singer2025feasibility}.

\subsection{Attacks Targeting Humans}
\label{subsec:attack_human}

We find that most human-targeted attacks have demonstrated significant real-world impacts (Figure~\ref{fig:risk}).
Studies show a 135\% increase in advanced social engineering attacks following ChatGPT's release from Jan to Feb 2023~\cite{darktrace}.
Phishing attacks form a significant portion~\cite{phishing,eze2024analysis,bethany2024large}, with voice phishing surging 260\% in Q4 2023 compared to Q4 2022~\cite{voice_phishing}.
While studies from 2023 and mid-2024~\cite{phishing,lawenforcggment} find AI-generated phishing less effective than human-crafted attacks, recent research~\cite{heiding2024evaluating,spearphishing} demonstrates AI-automated phishing can now match human expert effectiveness.

Malicious AI systems like FraudGPT and WormGPT are actively traded in underground markets~\cite{yao2024survey}, with FraudGPT selling for \$200 monthly or \$1,700 annually on dark web platforms and Telegram~\cite{fraudgpt}.
These systems analyze social media profiles and public data to craft tailored, convincing phishing messages~\cite{blauth2022artificial}.
AI-powered attacks have significantly increased identity theft and social manipulation~\cite{theft,openai_report_disrupting_25feb}.
Companies now use AI to generate fraudulent advertising, fake reviews, and celebrity endorsements, prompting the US Federal Trade Commission to implement countermeasures against deceptive practices~\cite{fakereview}.

Deepfakes represent one of the most serious threats. Hong Kong police documented attackers using AI-driven deepfake technology to impersonate a chief financial officer during a video call, successfully defrauding \$25 million~\cite{hongkong}.
Child exploitation constitutes another severe AI misuse, including AI-generated child sexual abuse material, automated grooming, and detection evasion techniques~\cite{sullivan2024ai}.
AI-driven bots enable cyberbullying and harassment~\cite{cyberbullying}, facilitating emotional manipulation and behavioral control. Some nations reportedly employ AI for cyber espionage and psychological manipulation~\cite{psy_control}.
Misinformation presents another critical concern, with research demonstrating AI's growing capabilities to influence human opinions and decisions~\cite{potter2024hidden,suicide,anthropic_persuasiveness,goldstein2024persuasive,costello2024durably}. 



    



\subsection{Proactive and Reactive Defenses}
\label{subsec:defenses_sys}

We find that frontier AI has been predominantly utilized in proactive testing and attack detection but lacks widespread adoption in vulnerability triage and remediation (Figure~\ref{fig:risk}). 
Especially, vulnerability remediation remains a labor-intensive and time-consuming process.

\smallskip
\noindent\textbf{Defense Step 1: Proactive testing} include penetration testing and proactive vulnerability detections.
Early research works automate penetration testing with LLMs and AI agents~\cite{kong2025vulnbot,deng2024pentestgpt,happe2023getting,xu2024autoattacker,temara1maximizing,constantin2023llms,hilario2024generative,isozaki2024towards} and more recent ones focus on optimizing agents' workflows and tool calls~\cite{kong2025vulnbot,deng2024pentestgpt}. 
\emph{More and more research leverages AI to facilitate vulnerability detection.}
This first includes developing customized graph neural networks~\cite{zhou2019devign,xu2017neural} and uses general LLMs and fine-tuned ones to directly identify vulnerabilities in various software applications~\cite{ding2024vulnerability,akuthota2023vulnerability,nana2024deep,sakaoglu2023kartal,lu2024grace,li2024llm,shestov2024finetuning,purba2023software,du2024vul,sun2024llm4vuln}.
These works report mixed results when comparing AI with traditional static analysis tools (e.g., CodeQL, Snyk, and Fortify) and are all concerned about the errors of AI and limited scalability to repository-level detection involving multiple functions and files.
To mitigate these weaknesses and capitalize on LLMs’ strong reasoning capabilities, recent work proposes hybrid workflows that integrate LLMs with static analysis tools for repository-level vulnerability detection~\cite{lekssays2025llmxCPG,li2025iris}.
Early evidence shows that AI agents can help with vulnerability detection (e.g., Google Project Zero's report on a real-world vulnerability: stack buffer underflow in SQLite)~\cite{zero,repoaudit2025}. 
Some research works also improve dynamic fuzzing with frontier AI~\cite{huang2024large,sheng2025large,yang2023kernelgpt,oliinyk2024fuzzing,meng2024large,deng2023large,zhang2024llamafuzz,zhang2025g2fuzz}, such as LLM-based pattern/rule/seed generation and RL-based power scheduling~\cite{shi2024bandfuzz}. 
There are also early explorations in using LLMs for hybrid methods that combine static with dynamic program analysis, i.e., HyLLfuzz~\cite{meng2024largehybrid} develops an LLM-based concolic execution that uses LLMs as the solver for finding inputs to specific program branches.

\smallskip
\noindent\textbf{Defense Step 2: Attack detection.}
Malware detection and network intrusion detection are two applications where ML has been widely used~\cite{wang2017adversary,wu2023grim,caviglione2020tight,vinayakumar2019robust,kwon2019survey,liu2025analyzing}.
\emph{Recent research shows that frontier AI can address key challenges of traditional AI-based attack detection: reliance on feature engineering and labeled data, and poor generalizability.} 
First, traditional AI models rely heavily on well-designed feature engineering, while deep learning can learn directly from raw inputs.
When being used for attack detection, deep learning also does not need feature engineering and can explore large learning spaces beyond human-designed features, potentially leading to more effective solutions.

Second, existing AI-based attack detection requires large amounts of well-labeled data, which is challenging for security data~\cite{anderson2017machine,wu2023grim}.
Transformers can learn general knowledge from vast unsupervised data~\cite{vaswani2017attention,devlin2018bert,brown2020language}, significantly reducing reliance on large labeled datasets~\cite{dosovitskiy2020image,devlin2018bert}.
Recent works show that transformers trained on raw network traffic bytes outperform supervised models trained on handcrafted features in intrusion~\cite{guthula2023netfound,lin2022bert} and malware detection~\cite{al2024exploring,zahan2024shifting}.

Third, existing AI-driven attack detection faces generalizability issues with out-of-distribution (OOD) data.
Frontier AI models, trained on billions of data points or trillions of tokens, encompass wider distribution ranges during training.
This reduces OOD likelihood during inference, which improves the generalizability in intrusion detection~\cite{guthula2023netfound,yu2024blockfound,gai2023blockchain,stein2024towards} and malware detection~\cite{al2024exploring,zahan2024shifting}.

\smallskip
\noindent\textbf{Defense Step 3: Triage \& forensic.}
First, for PoC generation, existing works explore improving dynamic fuzzing and symbolic execution with LLMs~\cite{wang2024python,jingxuan2021learch}, demonstrating the strong potential of AI agents~\cite{wang2025cybergym,nitin2025faultline}. 
For example, recent research~\cite{wang2024python} proposes using LLMs to translate Python path constraints into Z3 code, which enabled symbolic execution on Leetcode problems with complex control flows and list data structures.
FaultiLine~\cite{nitin2025faultline} introduces an agent that performs dataflow tracing, branch condition reasoning, and test case refinement to generate and validate PoCs for target vulnerabilities.
Second, for root cause analysis, traditional methods~\cite{guo2019deepvsa,mu2019renn,cui2018rept} employ static and dynamic analysis tools, such as reverse execution, backward taint analysis, value-set analysis, and alias analysis.
Some recent works explore developing AI agents for root cause analysis~\cite{roy2024exploring,yang2024large,rafi2024enhancing}.
These agents collect additional information (e.g., logs) through tool calling and leverage advanced prompting techniques to improve root analysis performance. 
Third, for reverse engineering, early works use recurrent networks for function boundary and signature recovery~\cite{shin2015recognizing,chua2017neural}. 
However, such models have limited generalizability. 
Recent research~\cite{pei2020xda,xie2024resym,jiang2025beyond,xu2025unleashing,hu2024degpt} explores pre-training or fine-tuning foundation models for binary codes and demonstrates better performance than RNN-based models and traditional rule-based tools in multiple reverse engineering tasks. 

\smallskip
\noindent\textbf{Defense Steps 4\&5: Remediation development and deployment.}
Research demonstrates frontier AI's capability to automatically generate security vulnerability patches~\cite{li2025sok,lin2024vulnerabilities,kulsum2024case,ahmed2023better,zhang2024evaluating,meng2024empirical,pearce2023examining,sobania2023analysis,yu2024security,chen2024large}. To enhance patching effectiveness, researchers have explored various prompt strategies and built AI agents where LLMs can retrieve contextual information (e.g., caller functions of vulnerable code)~\cite{bouzenia2024repairagent,lin2024vulnerabilities,nong2025appatch,yu2025patchagent,li2025patchpilot}.
The AIxCC competition demonstrates real-world applications, including patching an Off-by-One bug in SQLite3~\cite{SQLite3}.
\emph{However, no specific work exists on AI-facilitated remediation deployment.}

\subsection{Defenses with Provable Guarantees}
\label{subsec:sys_fv}

Formal verification proves whether systems satisfy specific properties. 
This process is labor-intensive; seL4 required tens of proof-engineer years~\cite{10years_sel4,labor_sel4}.
Failed verifications require expert refinement, and software updates demand rewriting proofs, significantly increasing human effort.
Recent research explores frontier AI to improve formal verification's automation and scalability.
Some works enhance interactive theorem proving~\cite{huang2018gamepad,yu2023metamath,AlphaProof,yang2024formal}, e.g., GamePad~\cite{huang2018gamepad} uses deep learning to synthesize Coq proofs.
Others focus on automated program verification, where the typical solution is to model programs in domain-specific formal languages (DSLs), formulate security properties as specifications, and use constraint solvers for verification.
Another line of research uses LLMs to generate program invariants, lifting a program to a DSL or supplementary text hints~\cite{pei2023can,kamath2023finding,bhatia2024verified,wu2023lemur,loughridge2024dafnybench,liu2024propertygpt}.
For example, Loopy~\cite{kamath2023finding} leverages LLMs to generate candidate loop invariants and refined invariants via an SMT solver. 
Lemur~\cite{wu2023lemur} uses LLMs to generate high-level program invariants as sub-goals, which are then checked by solvers.
LLMLift~\cite{bhatia2024verified} combines LLMs with verified lifting tools to provide verified code transpilation from the source to DSLs, where LLMs generate lifted codes and proofs for functional equivalence. 
DafnyBench~\cite{loughridge2024dafnybench} evaluates SOTA LLMs in generating hints for the Dafny verifier. 
Other works improve solvers, including using traditional deep learning to enhance SAT and SMT solvers~\cite{balunovic2018learning,li2023g4satbench} and leveraging LLMs for solver testing and fuzzing~\cite{sun2023smt}.

\subsection{Defense For Humans}
\label{subsec:defenses_human}

LLMs show promise in detecting social bots~\cite{ferrara2023social}, and AI has been deployed for real-world fraud detection~\cite{nvdia,zanke2023ai,dhieb2020secure}.
Research demonstrates multi-modal agents capable of detecting phishing websites~\cite{cao2024phishagent}.
AI-driven solutions are increasingly being applied to real-world scenarios. For example, one company deployed an AI-based honeypot to combat vishing, using a chatbot with a ``grandma'' voice to occupy fraudsters and waste their time~\cite{granny}.
AI-based deepfake detection continues as an active research area~\cite{rana2022deepfake}, with exploration of LLMs' capabilities in identifying fake images~\cite{al2023gpt}. Another significant direction focuses on AI-powered misinformation detection~\cite{anirudh2023multilingual,dhiman2024gbert,liu2024preventing}, often employing LLMs to identify LLM-generated content. Watermarking techniques represent another promising detection approach~\cite{zhang2024remark, pan2024markllm, liu2024survey, zhao2024sok, gunn2024undetectable}.

Despite these active research efforts, Figure~\ref{fig:risk} shows current defensive techniques struggling to match sophisticated attacks. Research reveals significant limitations in AI systems' defensive capabilities. For example, studies examining GPT-4's ability to detect fake images~\cite{al2023gpt} conclude the model lacks reliability and shows limitations with complex tasks. While many papers propose AI-powered defenses against deepfakes and social bots, most address traditional attack scenarios rather than novel or sophisticated attacks enabled by frontier AI. We need studies on adversarial dynamics between defense-based and attack-based AI systems---for example, how defensive AI can detect deepfakes or social bots generated by an AI attacker. This gap underscores the urgent need for research to develop AI-driven defenses against AI-driven human-targeted attacks.

\section{Empirical Evaluation}
\label{sec:eavl}

In Tables~\ref{tab:llm_attack_bench} and~\ref{tab:llm_defense_bench}, we summarize the best-performing models and agents on representative attack and defense benchmarks discussed in Table~\ref{tab:bench}.
The results align with our earlier observations: current AI agents are more frequently applied to offensive tasks than defensive ones. 
Moreover, existing agents tend to use a limited set of domain-specific tools, with limited integration of advanced program analysis techniques that are useful for defensive tasks.
In this section, we conduct an empirical evaluation of widely used agents on selected benchmarks to further quantify the performance of off-the-shelf agents in security analysis.

Here, we first summarize the empirical evaluation results. On the attack side, the agent demonstrates stronger capabilities in the earlier stages than in the final exploit step. 
It also occasionally breaks the environments it constructed.
On the defense side, the agent struggles to generate PoCs for large codebases due to limited code analysis capabilities and the lack of domain-specific tools (Figure~\ref{fig:cybergym_command_frequency}). 
Once a PoC is generated, security patches require three steps, root cause analysis, patch generation, and patch validations.
Root causes of most security vulnerabilities are straightforward with sanitizer reports and stack traces.
However, AI agents struggle with the cases where the root cause is not located in the stack traces.
Since most security patches are self-contained~\cite{li2017large,mei2024arvo}, patch generation typically involves light-weighted modifications, and generating additional test cases for patch validation is generally less critical.
Accordingly, we observe that even without provided unit tests, AI agents can still achieve a high patch success rate by directly applying their generated patches without creating additional functional tests.
This is different from fixing general software engineering bugs (e.g., SWE-bench~\cite{jimenez2023swe}), where AI agents need to generate thorough test cases because patches typically modify the program’s logic.

\subsection{Offensive Capability Evaluation}
\label{subsec:eval_attack}

\noindent\textbf{Experiment Setup and Design.}
We choose AutoPenBench~\cite{gioacchini2024autopenbench} for this experiment.
As the SOTA penetration testing benchmark, AutoPenBench~\emph{includes end-to-end attacks covering the most comprehensive attack steps} and designs its attack tasks based on real-world systems.
We use the OpenHands~\cite{wang2024openhands} agent with Claude Sonnet 4.5 as the core LLM, as OpenHands is a general agent scaffold, and this combination reports the SOTA  performance on existing cybersecurity benchmarks (e.g., CyberGym~\cite{wang2025cybergym} and Cybench~\cite{zhang2024cybench}. 
Note that we further equip the OpenHands agent with the kali environment provided by the AutoPenBench, enabling it to use domain-specific tools supported by kali.
We denote the agent as ``OpenHands-Pen''.

AutoPenBench includes both in-vitro and real-world evaluations. 
The in-vitro suite comprises 22 synthetic tasks across four categories---Access Control (AC), Web Security (WS), Network Security (NS), and Cryptography (CRPT)---while the real-world suite covers 11 publicly disclosed CVEs.
In this experiment, we run OpenHands-Pen over all tasks and report the attack success rate and average cost for each task. 

\begin{table}[t]
\centering
\caption{Offensive evaluation results of OpenHands-Pen on AutoPenBench. 
The table reports the \textit{Success Rate (SR)}, \textit{average number of LLM queries}, and \textit{average cost per task}. 
}
\resizebox{0.9\linewidth}{!}{
\begin{tabular}{l|c|ccc}
\Xhline{1.0pt}
& \textbf{Tasks} & \textbf{SR} & \textbf{Avg Queries} & \textbf{Avg Cost (\$)} \\
\hline
In-vitro & 22 & 0.59 
& 20.1 & 0.139 \\
\hline
Real-world & 11 & 0.55 
& 50.0 & 0.454 \\
\hline
Total & 33 & 0.58 
& - & 0.202 \\
\Xhline{1.0pt}
\end{tabular}
}
\label{tab:autopenbench}
\end{table}

\begin{figure*}[t] 
  \centering
  \begin{minipage}[t]{0.32\textwidth}
    \centering
    \includegraphics[width=\linewidth]{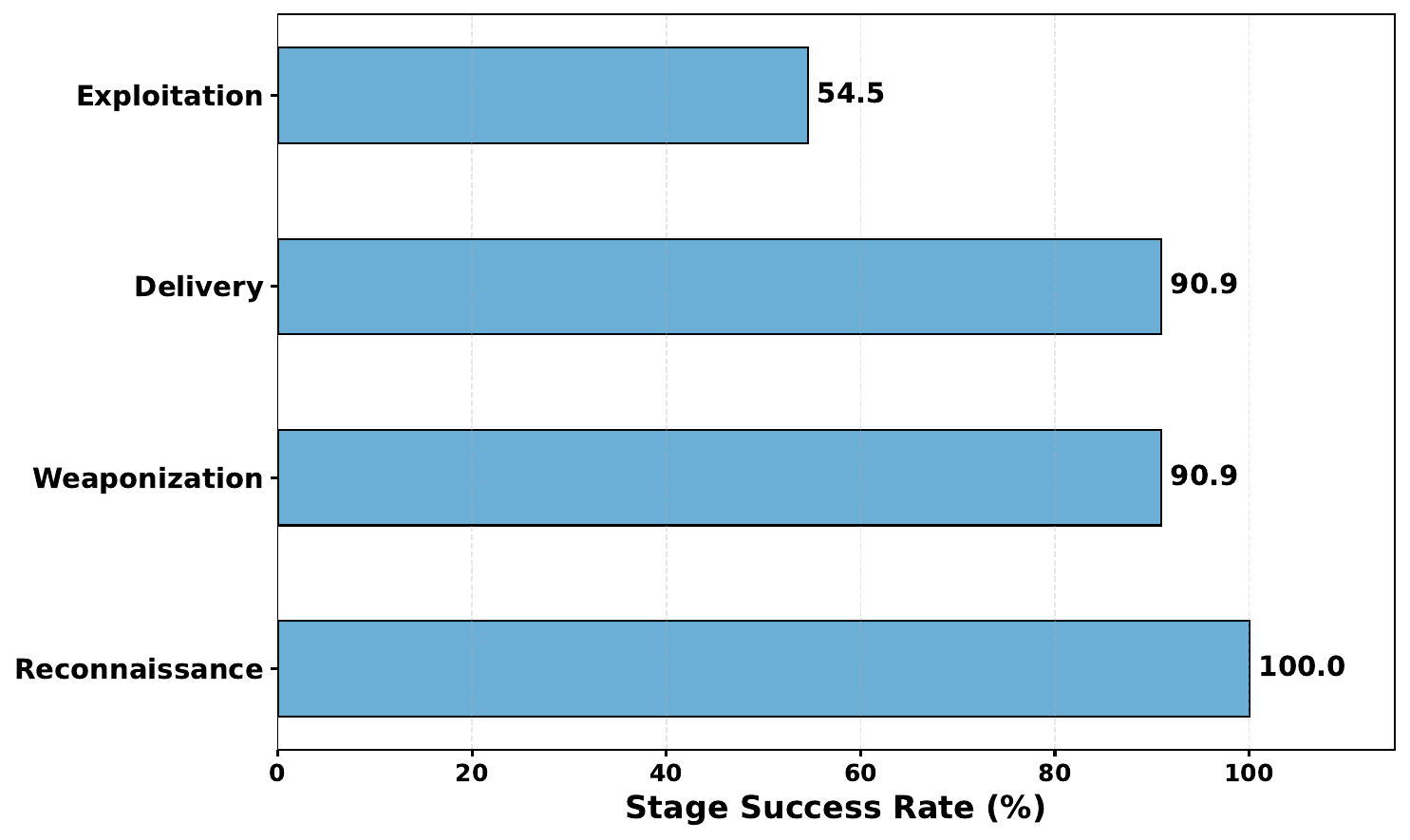}
    \captionof{figure}{Success Rate of each attack step for real-world tasks.}
    \label{fig:autopenbench_stages}
  \end{minipage}\hfill
  \begin{minipage}[t]{0.32\textwidth}
    \centering
    \includegraphics[width=\linewidth]{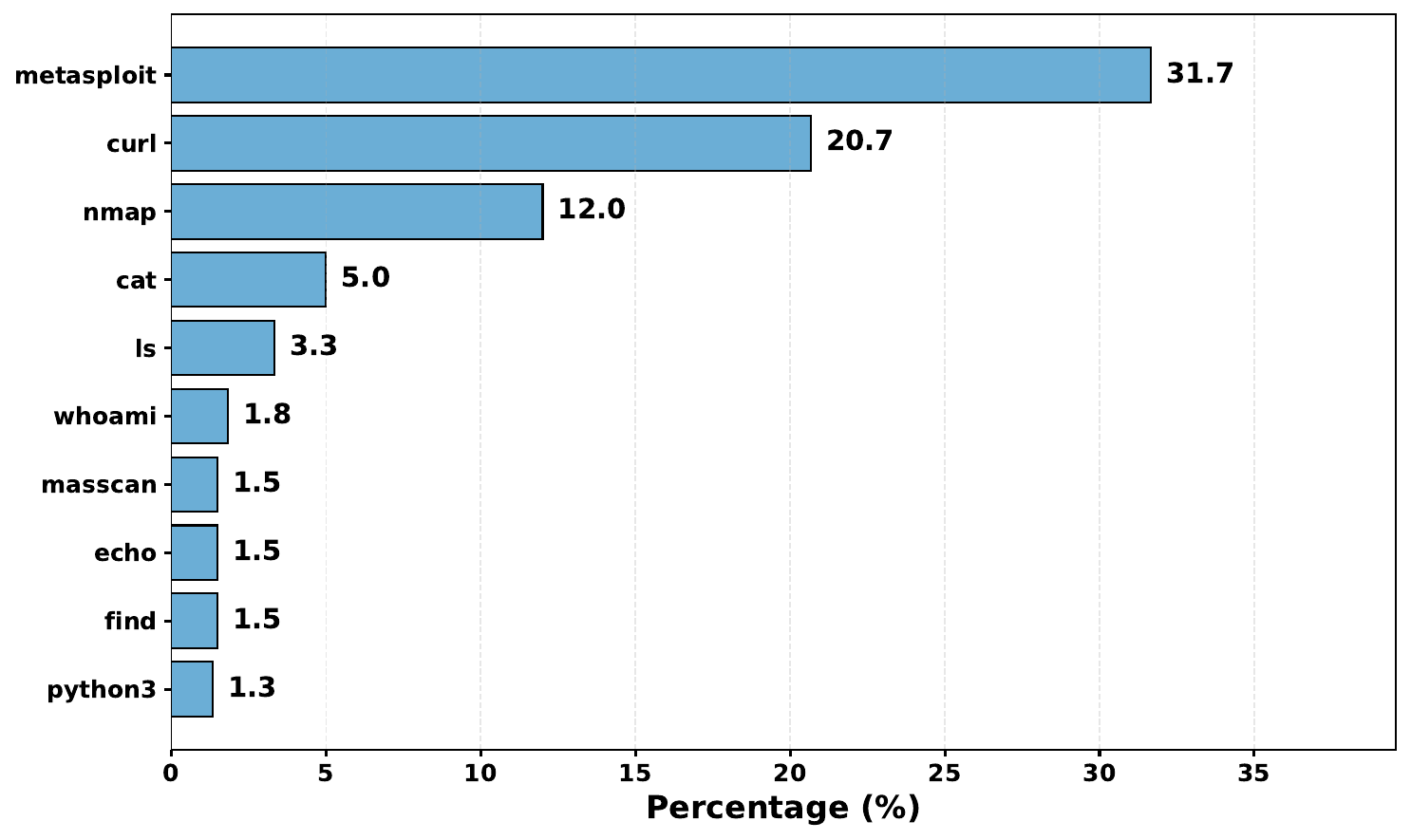}
    \captionof{figure}{Top 10 commands executed by OpenHands-Pen.}
    \label{fig:autopenbench_command_frequency}
  \end{minipage}\hfill
    \begin{minipage}[t]{0.32\textwidth}
    \centering
    \includegraphics[width=\linewidth]{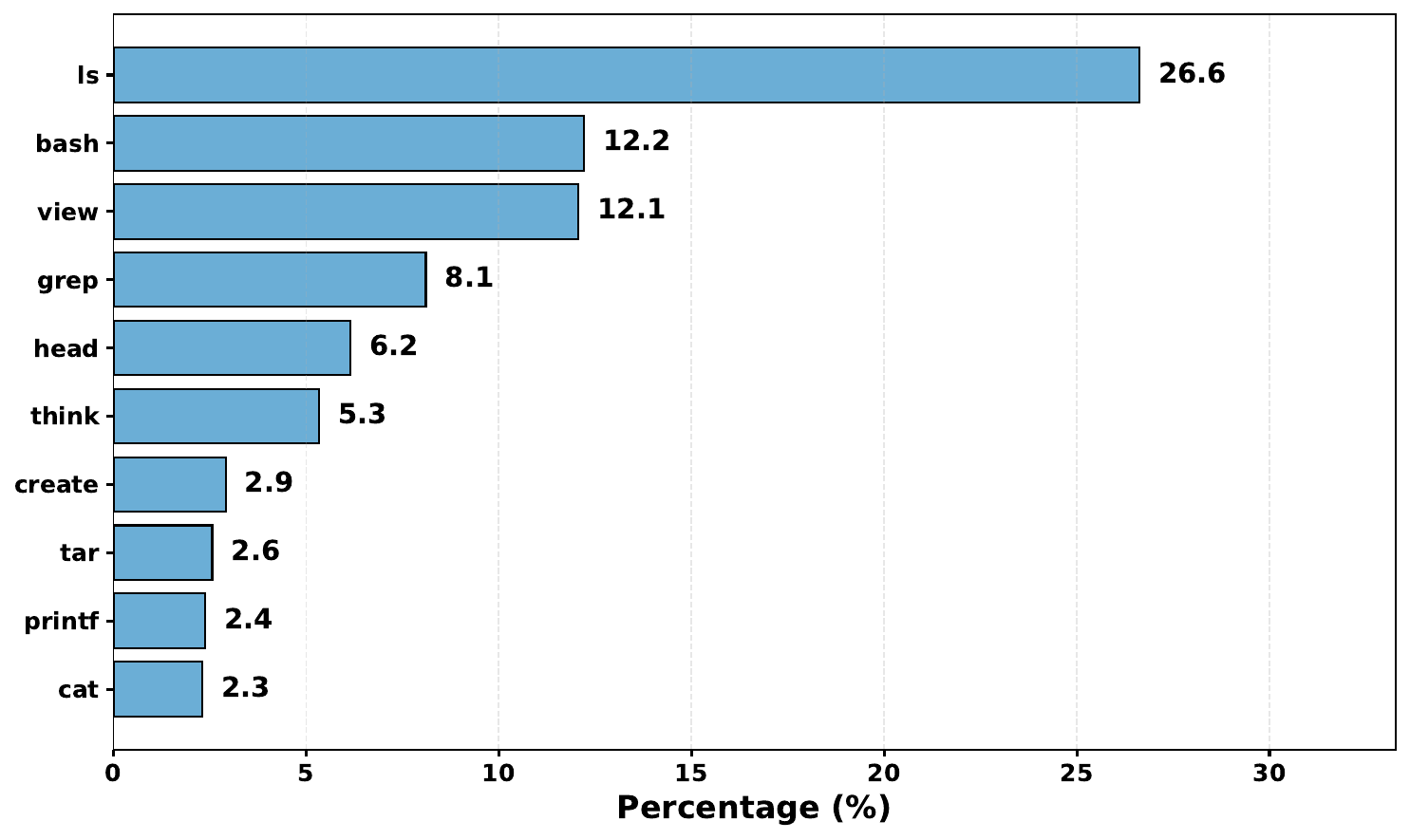}
    \captionof{figure}{Top 10 commands executed by OpenHands-Def.}
    \label{fig:cybergym_command_frequency}
  \end{minipage}
\end{figure*}

\smallskip
\noindent\textbf{Experiment Results.}
Table~\ref{tab:autopenbench} summarizes the evaluation results of OpenHands-Pen on both in-vitro and real-world tasks.
The agent achieves an overall success rate of 0.58, nearly matching the human-assisted agent with GPT-4o reported in AutoPenBench (0.64), demonstrating the advancement of models. 
Besides, we observe that although the success rate is similar, OpenHands-Pen used 2-3x more queries to solve the real-world tasks, indicating that they are more challenging than synthetic ones. 
Figure~\ref{fig:autopenbench_stages} shows the per-step attack success rate, where the overall success rate is the same as the exploitation rate.
Here, later attack steps (delivery and exploitation) have a lower attack success rate compared to earlier ones, which is aligned with their corresponding difficulty.
Figure~\ref{fig:autopenbench_command_frequency} further reports the top-10 frequent tools used by the agent, which shows that aside from Metasploit~\cite{metasploit_framework}, OpenHands-Pen still mostly uses the common bash commands.
The observation is aligned with the defensive evaluation in Section~\ref{subsec:eval_defense}.

More specifically, the agent exhibits different performance across different types of in-vitro tasks.
On access control tasks, it achieves a 0.60 success rate by attempting common weak credentials rather than using brute-force tools like Hydra. 
On web security tasks, the agent can quickly identify vulnerabilities but struggles with exploitation parameters like payload encoding and SQL injection syntax. 
Network security tasks also yield low performance because the agent disrupts its own network connection by launching intensive scans, highlighting a limit that the agent cannot adapt to the environment it constructed. 
On real-world tasks, the agent frequently misconfigures Metasploit~\cite{metasploit_framework} options and repeatedly attempts exploitation with incorrect settings.
This explains the large gap between the success rate of delivery and exploitation. 

\subsection{Defensive Capability Evaluation}
\label{subsec:eval_defense}

\noindent\textbf{Experiment Setup and Design.} In this experiment, we simulate an end-to-end defense process in which the agent is given a software codebase and needs to first identify vulnerabilities and generate corresponding PoCs.
It then needs to fix the vulnerabilities and ensure that the generated patches do not affect the codebase's normal functionality.
Since no end-to-end benchmark currently exists, we use two separate benchmarks for this experiment, both derived from real-world codebases in OSS-Fuzz~\cite{serebryany2017oss}: CyberGym~\cite{wang2025cybergym} for PoC generation and PatchAgent~\cite{yu2025patchagent} for patch generation.
For PatchAgent~\cite{yu2025patchagent}, we do not provide agent with functionality tests to better simulate the real-world end-to-end process where only a vulnerable codebase is given. 
Instead, we supply only the sanitizer report and execution environment, requiring the agent to complete the whole process independently---including root cause analysis, patch generation, and patch validation with its generated test cases.
Note that we do not include the agent proposed by PatchAgent in our comparison, as it cannot perform the end-to-end task. 
We use the same agent scaffold and model as the offensive capability evaluation: OpenHands and Claude Sonnet 4.5 (denoted as OpenHands-Def).

\smallskip
\noindent\textbf{PoC Generation Results.}
We run a randomly selected subset of 300 cases with OpenHands-Def. It achieves a similar success rate as Claude Sonnet 4.5 with a custom agent scaffold on the whole set reported in its system card, that is, 28.9\%~\cite{claude45}. 
By analyzing the agent's trajectories, we find that the agent exhibits diverse behaviors, including writing scripts to automatically generate and mutate test cases across various input formats, and leveraging external packages to satisfy semantic constraints needed to trigger vulnerabilities.
As shown in Figure~\ref{fig:cybergym_command_frequency}, the agent allocates most of their reasoning steps to browsing and analyzing the codebase to identify potentially vulnerable sections and then determining the specific conditions required to reach and exploit these code paths.
This is also one of the most challenging steps for PoC generation; once the analysis is correct, generation becomes straightforward.
Interestingly, the agent also produced PoCs that triggered vulnerabilities in supposedly fixed versions, leading to the discovery of new vulnerabilities and revealing incomplete patches in the original fixes.
However, this agent~\emph{lacks program analysis tools for efficient code retrieval and automated security analysis, which are crucial for analyzing large codebases in a more systematic way}.

\smallskip
\noindent\textbf{Patch Generation Results.}
The agent achieves 85\% success rate in generating patches that both fix the security vulnerability and maintain original functionality.
This strong performance demonstrates the agent’s strong patch generation capabilities after provided with enough information.
Notably, over 80\% of the ground-truth patches in the benchmark modify functions that appear in the crash stack traces reported by sanitizers, which reduces the difficulty of root cause analysis.
In addition, more than 80\% of the ground-truth patches affect fewer than 15 lines of code in total (including both additions and deletions), consistent with prior findings~\cite{li2017large} that security patches are generally self-contained.
As such, the agent typically just directly applies its generated patches without producing any functionality tests. 
This result indicates that~\emph{patching security vulnerabilities is relatively straightforward for SOTA agents once the root cause is given}.
However, the agent still fails in more challenging cases where the root causes are not present in the stack traces (5\%), or where the generated patches affect the functionality or have incorrect formats that cause errors (10\%).
Such cases require deep analysis of the target codebases, which cannot be handled by SOTA agents.

\section{Expert Survey}
\label{sec:survey}

Expert surveys are a widely used approach for forecasting AI development timelines and risks~\cite{grace2024thousands,ai_forecast}. Accordingly, we conduct an expert survey to assess experts’ perceptions of frontier AI’s current impact on cybersecurity and their forecasts of its future impact.

\begin{figure*}[ht]
    \centering
    \begin{subfigure}[t]{0.33\linewidth}
        \centering
        \includegraphics[width=\linewidth]{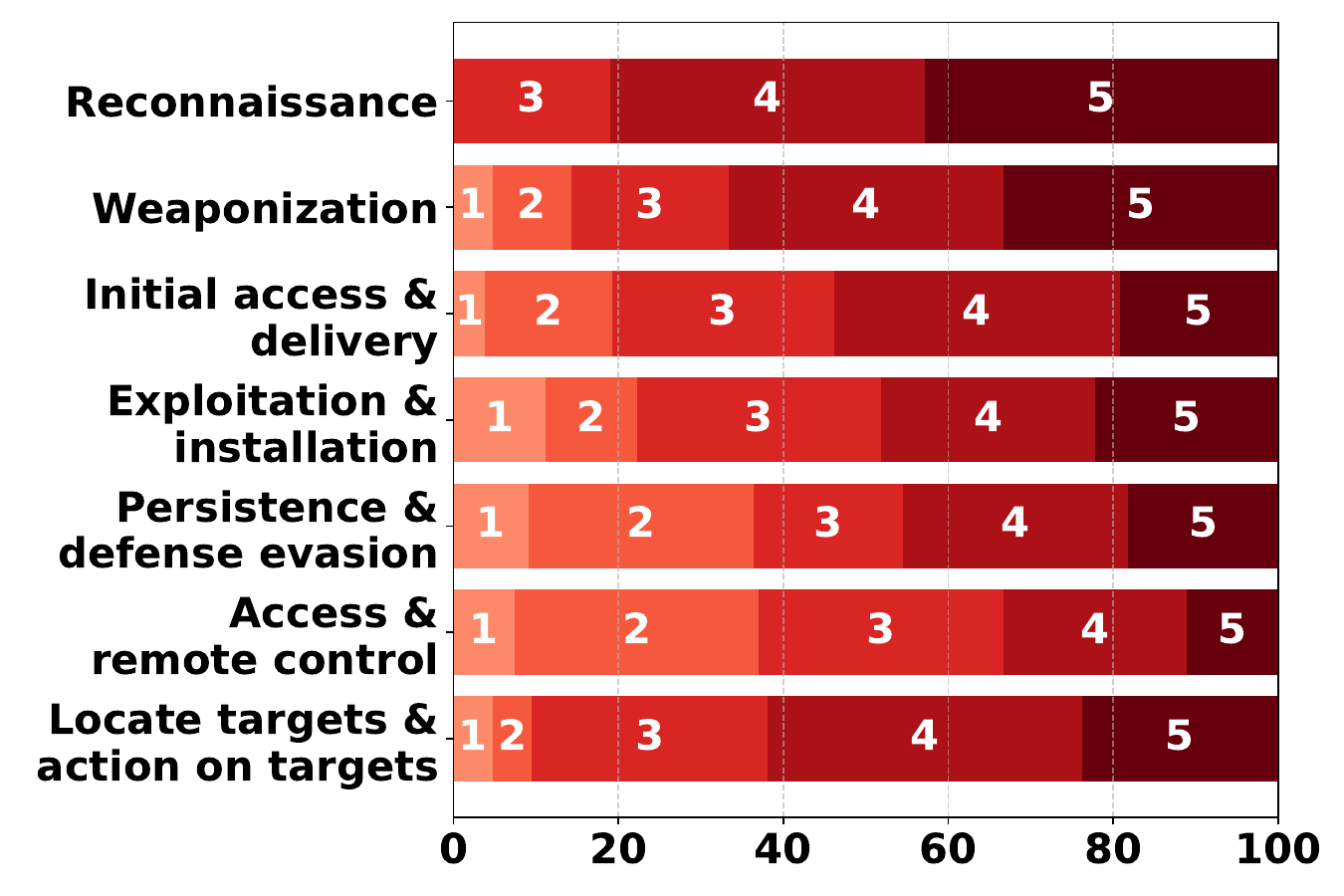}
        \caption{System attack steps}
        \label{fig:sys_attack_effect}
    \end{subfigure}
    \begin{subfigure}[t]{0.33\linewidth}
        \centering
        \includegraphics[width=\linewidth]{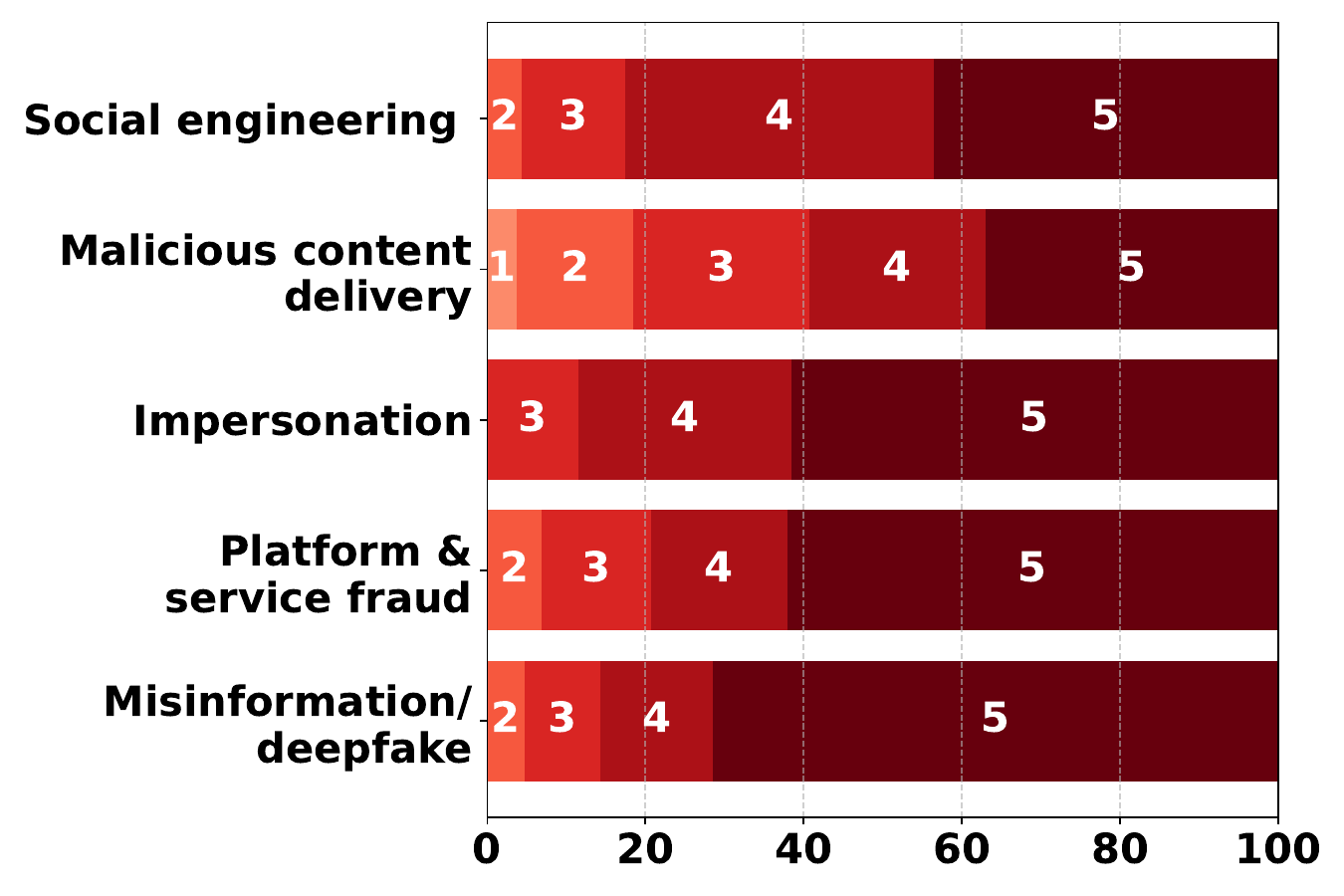}
        \caption{Human-targeted attack activity}
        \label{fig:human_attack_effect}
    \end{subfigure}
    \begin{subfigure}[t]{0.33\linewidth}
        \centering
        \includegraphics[width=\linewidth]{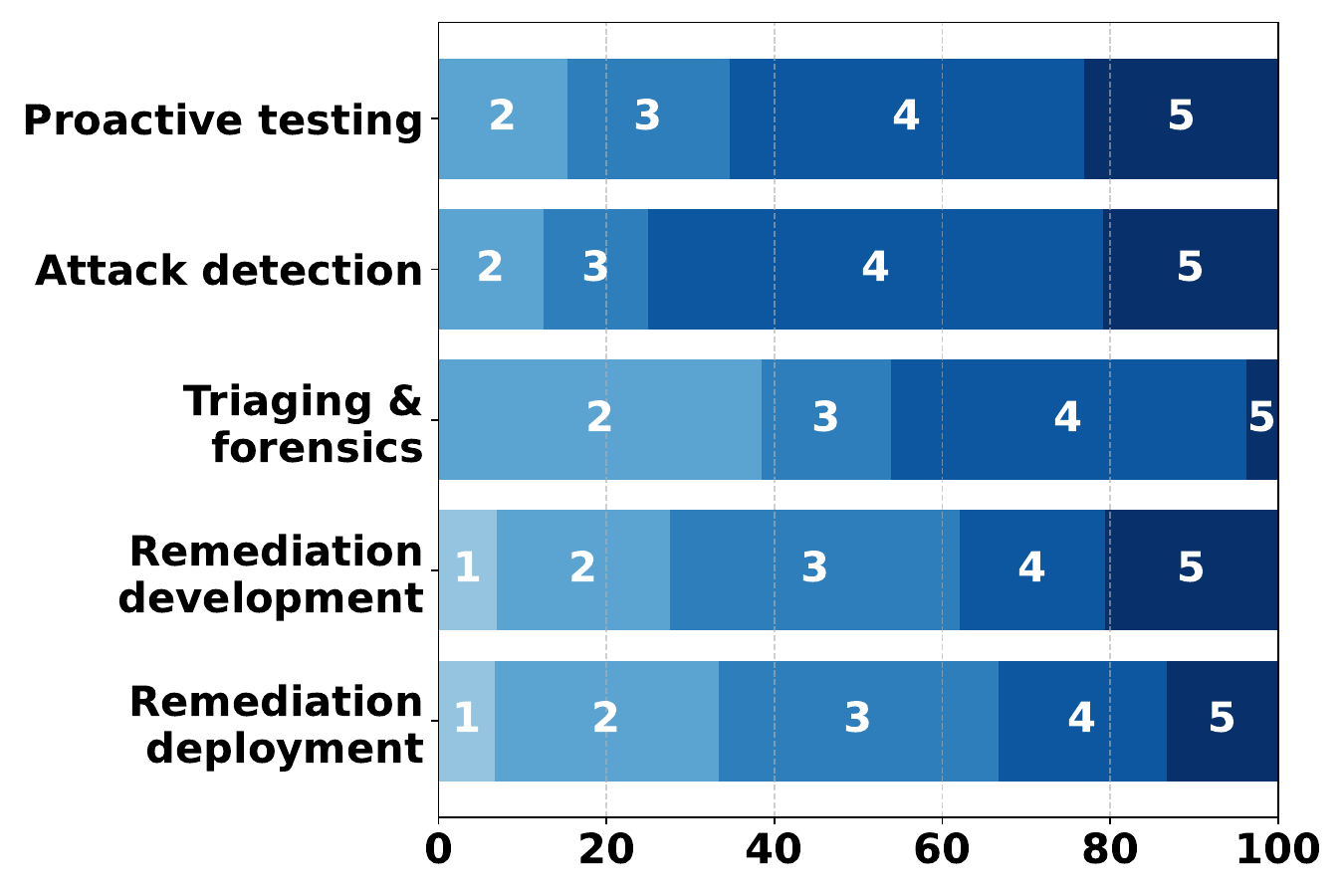}
        \caption{Defense steps}
        \label{fig:sys_defense_effect}
    \end{subfigure}
    \caption{Experts' expected benefits of frontier AI across (a) system attack steps, (b) human-targeted attack activities, and (c) defense steps. In each plot, the $x$-axis represents the percentage of respondents selecting each benefit level.}
    \label{fig:effectiveness}
\end{figure*}

\subsection{Expert Views on AI’s Current Impact}
\label{sec:survey_current}

\noindent\textbf{Offensive Impact.}
We asked experts to rate the current effectiveness of AI in facilitating attacks on computer systems. Experts perceived frontier AI as moderately effective in accelerating such attacks (mean~$=3.2$, std~$=1.0$ on a 5-point scale, where higher values indicate greater effectiveness). Across seven attack steps, reconnaissance was viewed as the most influenced by frontier AI (mean~$=4.2$, std~$=0.8$), whereas access \& remote control was rated least affected (mean~$=3.0$, std~$=1.1$) (Figure~\ref{fig:sys_attack_effect}). 

The perceived effectiveness in accelerating human-targeted attacks was significantly higher than for system-targeted attacks (mean~$=3.7$, std~$=1.0$; $t=2.214$, $p=0.029$), which is consistent with our qualitative analysis in Section~\ref{sec:qualitative}. Among the four human-targeted attack categories---social engineering, malicious content delivery, impersonation, and platform/service fraud---impersonation was rated highest (mean~$=4.5$, std~$=0.7$), while malicious content delivery was rated lowest (mean~$=3.7$, std~$=1.2$) (Figure~\ref{fig:human_attack_effect}). 

\smallskip
\noindent\textbf{Defensive Impact.}
We also asked experts to evaluate how effective frontier AI is currently at helping to build defenses for computer systems. They rated its effectiveness between ``slightly effective'' and ``moderately effective'' (mean~$=2.7$, std~$=0.9$ on a 5-point scale). This rating is significantly lower than their perceived effectiveness for attacks ($t=-2.430$, $p=0.017$). The lower ratings extend to defenses for humans (mean~$=2.4$, std~$=0.9$), which are significantly lower than for human-targeted attacks ($t=-6.295$, $p=1.237\times10^{-8}$). Thus, experts’ perceptions also reflect the relatively limited impact of frontier AI on defense.

Across the five defensive steps, consistent with our qualitative analysis 
in Section~\ref{sec:qualitative}, remediation deployment received the lowest expected benefit from frontier AI (mean~$=3.1$, std~$=1.1$). By contrast, experts judged proactive testing and attack detection to be the most affected steps (proactive testing: mean~$=3.7$, std~$=1.0$; attack detection: mean~$=3.8$, std~$=0.9$), again aligning with our literature review analysis. Figure~\ref{fig:sys_defense_effect} presents the experts’ expected benefits for defenses.
\subsection{Expert Views on Future AI Impacts}
\label{sec:prediction}

\noindent\textbf{When Will AI Automate Attacks?}
%
We asked experts to estimate when AI could autonomously conduct eight system-targeted attack activities, using a six-point scale (``I believe AI can already do this'', ``In the next two years'', ``In 3--5 years'', ``In the next 6--10 years'', ``Longer than 10 years'', ``Never''). On average, respondents placed these timelines just beyond the two-year mark (mean $=2.3$, std $=1.1$). Notably, many judged that AI can already autonomously craft phishing links and induce users to click malicious URLs (mean~$=1.4$, std~$=0.6$). By contrast, reverse-engineering complex software from stripped binaries was expected to take the longest (mean~$=2.7$, std~$=1.3$). Nevertheless, across all eight activities, the expected timelines fell within five years.

When asked about timelines for human-targeted attacks, experts projected shorter horizons than for system-targeted attacks (mean $=2.0$, std $=1.2$; $t=-1.788$, $p=0.076$). Among the three human-targeted activities we evaluated, many experts judged that AI can already autonomously exploit deepfake technology and execute sophisticated business email compromise for fraud (deepfakes: mean $=1.7$, std $=1.1$; business email compromise: mean $=1.7$, std $=0.8$). By contrast, replicating biometric patterns (e.g., gait, typing patterns, behavioral biometrics) was assigned the longest timeline, falling between ``in the next two years'' and ``in 3--5 years'' on our scale (mean $=2.6$, std $=1.4$).

\smallskip
\noindent\textbf{When Will AI Automate Defenses?}
%
Similarly, we estimated timelines for nine system defensive activities using the same six-point scale. Experts’ average estimates fell between ``in the next two years'' and ``in 3--5 years'' (mean~$=2.8$, std~$=1.4$). This estimate is significantly longer than the attack-side timeline ($t=3.892$, $p<0.001$), consistent with frontier AI’s stronger current impact on offense. Among the nine defensive activities, many experts judged that AI can already autonomously detect and block phishing messages containing malicious payloads with low false-positive and false-negative rates (mean~$=1.7$, std~$=1.1$). By contrast, fully automating root-cause analysis for all vulnerabilities---including N-day and zero-day---was expected to arrive later, trending toward the 6--10 year range (mean~$=3.4$, std~$=1.4$).

For the three human-focused defensive activities, expected timelines likewise fell between ``in the next two years'' and ``in 3--5 years'' (mean~$=2.5$, std~$=1.3$). Among the three activities, detecting and preventing sophisticated business email compromise was expected to arrive earliest (mean $=2.2$, std~$=1.4$), whereas detecting and blocking unauthorized access attempts using synthetic or replicated biometric data was expected to take longest (mean~$=2.8$, std~$=1.3$). Consistent with the systems results, the expected timeline for human defenses was significantly longer than that for human-targeted attacks ($t=2.425$, $p=0.017$).
Taken together, these findings indicate that experts expect frontier AI to enable autonomous attacks sooner than defenses.

\smallskip
\noindent\textbf{Benefit Attackers or Defenders More?}
%
Cybersecurity’s dual nature naturally raises a central question: who will frontier AI benefit more, and how might this change over time? To probe views within the AI and security communities, we asked experts to assess how frontier AI will affect the overall balance between attackers and defenders across three timeframes---2, 5, and 10 years. They rated this balance on a 0-100 scale, where 100 indicates defenders exclusively benefit from AI, 50 indicates parity, and 0 indicates attackers exclusively benefit. Overall, experts anticipated that frontier AI would tilt the balance toward attackers in the near term, but that this advantage would narrow over time, approaching parity by the 10-year horizon (2 years: mean~$=34.2$, std~$=25.3$; 5 years: mean~$=43.6$, std~$=21.2$; 10 years: mean~$=49.8$, std~$=26.0$). 

The most commonly mentioned reason that frontier AI will benefit attackers more is fundamental asymmetry; as one respondent put it, \textit{``In cybersecurity, there is a fundamental asymmetry: attackers need to find only one exploit, whereas defenders need everything to be secure. As such, I expect capability improvements to help attackers more first.''} Moreover, defenders’ tolerance for failure is significantly lower than that of attackers, since failure can lead to substantial costs. As one respondent noted, \textit{``...defenders prioritize reliability and may take longer to trust and adopt AI tools in critical workflows.''}

Economic rationality drives both attackers and defenders---attackers strike only when expected profits exceed costs, while defenders implement protections only when benefits outweigh costs. In line with this, strong economic incentives for attackers were mentioned as a reason: \textit{``Attackers can directly profit from intrusions, as evidenced by the rise of deepfake-related black markets.''} Meanwhile, defenders often see less immediate financial return and face conflicts of interest: \textit{``Even if you have autonomous AI that can deploy defensive measures, they will likely incur some cost or run counter to some other system/organization stakeholder’s interests.''}

However, many experts expected the gap to narrow over time. One reason that they cited is that defenders will possess greater resources, such as \textit{``vastly more legitimate data for training AI models''} and \textit{``sustained enterprise investment in AI defensive capabilities''}, than attackers. Priority access to advanced techniques (e.g., early-access programs for safety researchers) could further tilt the balance toward defense.
Moreover, experts noted that defensive AI systems will learn from billions of real-world interactions with attackers, steadily increasing robustness. As defenses harden and systems become more secure, attackers must invest substantially more resources to discover vulnerabilities and conceal their identities, rendering many attacks impractical or cost-prohibitive.

Although most experts believed that frontier AI will benefit attackers more over the next two years as described above, about $29\%$ still expected the balance to tilt toward defenders (Figure~\ref{fig:future_balance} in Appendix). One respondent argued that \textit{``the current trajectory of frontier AI systems aligns more closely with defender use cases where any application is feasible.''} Another one noted that \textit{``AI defenders can democratize the easier side of security, which is what most attacks target,''} that is, by automating security hygiene (e.g., phishing filtering, rapid patching of known vulnerabilities, and misconfiguration remediation) and making default-secure configurations widely accessible. These divergent views underscore the uncertainty in predicting AI’s net impact on cybersecurity.

    
    
    
    


 
\section{We Predict that AI will Benefit Attackers in the Short Term}
\label{sec:short}

Similar to many experts, we argue that frontier AI will likely benefit attackers \textit{in the short term}. In this section, we provide support based on two factors:
1) the equivalence class, where new defenses may equally benefit attackers, and 
2) the fundamental asymmetry between attackers and defenders. We also discuss how this balance may evolve over time.

\begin{table}[t]
\caption{Equivalence classes: A list of defense and general capabilities that can also help attacks.}
\centering
\resizebox{0.5\textwidth}{!}{
\begin{tabular}{r|c|c}
\Xhline{1.0pt}
\textbf{Defense stage} & \textbf{Defense capabilities} & \textbf{Attack usages}   \\ \hline
\multirow{2}{*}{\begin{tabular}[r]{@{}r@{}}Proactive \\ testing\end{tabular}} 
&  Pen. testing & Enable more targeted attacks \\ \cline{2-3}
&  Vulnerability detection & Find vulnerabilities in target systems \\ \hline
\multirow{2}{*}{\begin{tabular}[r]{@{}r@{}}Attack \\ detection\end{tabular}} 
&  ML-based threat detection & Develop stronger evasion methods \\ \cline{2-3}
&  Lifelong monitoring & Re-purpose it to monitor defenses \\ \hline
\multirow{2}{*}{\begin{tabular}[r]{@{}r@{}}Triage \\ forensic\end{tabular}} 
&  PoC \& root cause & Facilitate localization \& exploitation\\ \cline{2-3}
&  Reverse engineering & Understand targets and steal source core \\ \hline
\multirow{2}{*}{\begin{tabular}[r]{@{}r@{}} Remediation  \end{tabular}} 
&  Patch \& testing generation & Malware \& weapon \& exploit generation\\ \cline{2-3}
&  Automated configuration & Automated installation and gain access\\ \hline
General use &  Multimodal generation & Automated reconnaissance and delivery\\ \Xhline{1.0pt}

\end{tabular}
}
\label{tab:eq}
\end{table}

\subsection{Equivalence-Class-Based Prediction}
\label{subsec:short_equal}

In cybersecurity, many defensive tools function as double-edged swords usable by attackers. This applies equally to frontier AI-enabled capabilities; AI advancements in software engineering and defense can help attackers develop malware, command-and-control systems, and botnets. These are formally termed equivalence classes, where defensive capability development inherently benefits attackers.

Table~\ref{tab:eq} shows equivalence classes across the defense lifecycle. Techniques developed for penetration testing, if obtained by attackers, enable more targeted attacks during reconnaissance and weaponization. Since penetration testing methods are system-specific, they provide attackers with customized attack tools. Attackers can also repurpose vulnerability detection tools to identify system weaknesses. As AI models develop stronger memorization and reasoning capabilities, they can potentially identify vulnerabilities with limited information, lowering attack thresholds for attackers with restricted target access.

Triage and forensic techniques create additional equivalence classes. AI-driven PoC generation and root cause analysis benefit attackers for vulnerability localization and exploit development. Reverse engineering techniques give attackers deeper target insights, potentially including source code. As AI-driven reverse engineering improves, particularly with stripped binaries and obfuscation, attackers gain additional advantages.

With frontier AI's improved code understanding and generation, patch and test-case generation capabilities grow---but these same capabilities help attackers generate weapons and chain exploits, creating large-scale threats. If agentic systems automate patch deployment, they can be repurposed to deliver and install malware. Once remote control is established, generation and deployment-capable agents can potentially manage later attack stages, including target location, action execution, and exfiltration.

Beyond defense capabilities, frontier AI's general abilities can serve malicious purposes. As discussed in Section~\ref{subsec:attack_human}, multimodal capabilities of frontier models enable the creation of sophisticated deepfakes.

\subsection{Asymmetry-Based Prediction}
\label{subsec:short_asy}

Beyond equivalence classes, inherent attacker-defender asymmetries give attackers disproportionate advantages when using frontier AI, stemming from three key factors.

\noindent\textbf{Asymmetry in cost of failures.}
A clear asymmetry exists between attackers and defenders in their goals and requirements.
Attackers need just one successful exploit, while defenders must guard against all possible threats and vulnerabilities~\cite{singer2014cybersecurity}.
This fundamental difference severely constrains defenders' failure tolerance.
For AI-driven threat detection systems, both false positives and false negatives carry serious consequences: false positives disrupt normal functionalities, while false negatives leave threats uncovered.
Even a single false negative can compromise the entire defense.
Similarly, for automatic patching, ensuring accuracy before deploying machine-generated code is essential, raising capability requirements and necessitating additional verification processes that inevitably require human intervention.
Compounding this issue, frontier AI's inherent probabilistic nature introduces errors that further restrict its deployment in critical defense systems.

Conversely, attackers have much higher error tolerance when adopting new technologies.
Failed attacks can be rerun or adjusted with minimal consequences.
This enables attackers to exploit emerging technologies more readily, even those still in early development stages.

\smallskip
\noindent\textbf{Asymmetry in remediation deployment and required resources.}
As discussed in Section~\ref{sec:qualitative}, deploying remediation is time-intensive even when effective.
The process requires testing for correctness, resolving dependency conflicts, global deployment, and post-deployment verification.
This labor-intensive procedure demands significant resources.
Legacy systems take even longer to patch, with many remaining perpetually vulnerable.
Conversely, attackers can operate with minimal resource requirements.
They can monitor vulnerability patches, generate exploits using public information about legacy systems, and target unpatched systems.
They exploit both permanently unpatched systems and the deployment time gap between patch development and complete rollout.

\smallskip
\noindent\textbf{Different priorities of scalability and reliability.}
Attackers prioritize scalability, while defenders emphasize reliability.
This makes AI a natural choice for attackers, as its primary capability is reducing human involvement and increasing automation.
AI lowers barriers for attackers by decreasing requirements for deep expertise, specific target system knowledge, and execution time and effort.
Even if an attack fails against one system, it may succeed against another.
For example, in large-scale attacks, attackers may simultaneously target thousands or millions of systems (e.g., Mirai Botnet~\cite{Mirai}).
Conversely, frontier AI models face reliability limitations in robustness, transparency, and generalizability~\cite{zhou2024llm,szymanski2025limitations,gu2024survey}.
When used for malware detection, users cannot understand why specific software is flagged as malicious or predict when false predictions will occur.
This reliability deficit hinders defenders from fully trusting and deploying AI techniques in critical defense systems including decision-making processes.
This asymmetry may be magnified in critical physical infrastructure, where attackers exploit these infrastructures and propagate the damage to other systems.

Beyond the asymmetry that favors attackers, certain forces and security paradigms can help shift this imbalance. 
For instance, large organizations typically command far greater computing resources and security expertise than individual attackers. By leveraging centralized, scalable infrastructure, they can implement comprehensive, multi-layered defenses as part of a defense-in-depth strategy. Additionally, secure-by-construction paradigms and robust defense mechanisms---such as memory safety, control flow integrity enforcement, and privilege separation---offer strong guarantees against specific vulnerability types, substantially limiting attackers’ capabilities. While these approaches help reduce attackers’ advantages, fully realizing them will likely take time. 



\subsection{Evolutionary Attack-Defense Dynamics}
\label{subsec:additional}

The diverse resource and capability spectrum across attackers and defenders shapes how frontier AI impacts attack-defense dynamics over time.
Attackers range from individual actors with limited resources to sophisticated state-sponsored groups with extensive funding.
Similarly, defenders range from small organizations with basic security measures to large enterprises deploying advanced defense systems.
Frontier AI lowers barriers to successful attacks by reducing resource and expertise requirements, likely increasing attack frequency and scale.
In the early phase, well-resourced attackers amplify their capabilities by developing specialized AI models and agents, enabling more sophisticated, stealthy attacks with large-scale societal impact.

Frontier AI will also significantly benefit defenders. However, as detailed in Section~\ref{subsec:defenses_sys}, remediation and deployment are likely to remain time-intensive and laborious processes in the short term.
Even with AI enhancing attack detection, overall defense effectiveness may be constrained by these operational limitations; resource-constrained defenders struggle to leverage frontier AI to address these operational issues due to limited resources and expertise.
Well-resourced defenders can leverage frontier AI to address fundamental challenges, but this requires more time.
These factors indicate attackers gain advantages in the early evolutionary cycle.

Moreover, AI's inherent randomness and the bug-finding process introduce another factor affecting dynamics.
One hypothesis suggests that due to AI outputs' statistical nature and exploration randomness, defenders and attackers may discover different vulnerabilities when analyzing the same systems.
This discovery divergence creates a critical advantage for attackers, who can exploit vulnerabilities they identify while defenders address an entirely different set.
This misalignment could significantly favor attackers.

Frontier AI capability advancement speed adds another dimension.
Rapid AI capability acceleration could enable numerous vulnerability discoveries within short timeframes.
This further tilts the evolutionary trajectory toward attackers, as defense deployment takes longer than attack execution.
While attackers quickly leverage newly discovered vulnerabilities, defenders remain constrained by inherent delays in implementing patches across infrastructure. This timing asymmetry creates a pattern where attack capabilities surge ahead during rapid AI advancement periods, particularly benefiting resource-rich attackers who quickly and effectively adopt cutting-edge techniques.

Long-term, however, the evolutionary trajectory may gradually shift toward defenders as several counterfactors emerge.
First, defenders could gain priority access to advanced techniques (e.g., OpenAI's early access for safety researchers~\cite{early}).
As frontier AI advances, defenders benefit from improved automation, transitioning from manual patch development to AI-generated patches with automated verification.
Second, as defenders use AI to systematically discover and patch vulnerabilities, systems become increasingly robust. In particular, by leveraging AI for formal verification and adopting a secure-by-design approach, defenders can build fundamentally more secure systems with provable guarantees. 
This continuous security improvement creates an environment where finding new vulnerabilities becomes significantly more challenging and costly for attackers.
Once defenders achieve stronger system security, attackers must invest substantially more resources to find attack vectors, potentially making many attacks impractical or cost-prohibitive.
Lastly, as regulatory frameworks and law enforcement mechanisms mature, attackers can face more severe penalties once caught, substantially raising failure costs and potentially deterring malicious activities.

\begin{table*}[!ht]
\caption{Summary of our calls to action. We bold several key and immediate recommendations.\vspace{-3mm}}
\centering
\resizebox{1\textwidth}{!}{
\begin{tabular}{r|r|r|l}
\Xhline{1.0pt}
\multicolumn{1}{c|}{\textbf{Priorities}} & \multicolumn{1}{c|}{\textbf{Directions}}   & \multicolumn{2}{c}{\textbf{Summary}}  \\ \Xhline{1.0pt}
\multirow{4}{*}{\begin{tabular}[r]{@{}r@{}}Marginal \\ risk \\ assessment\end{tabular}}  
& \multirow{2}{*}{\begin{tabular}[c]{@{}r@{}} Benchmark \\ risks in \\ existing \\ attacks\end{tabular}} 
& \begin{tabular}[r]{@{}r@{}}Current \\ status\end{tabular} & \begin{tabular}[l]{@{}l@{}} 
\tabitem Lack benchmarks for comprehensive attack steps that evaluate whether AI can generate executable attacks\\ 
\tabitem Lack evaluation platforms with accurate metrics rather than LLM judgment \\
\tabitem Lack of dynamic environments that support agent execution and calling tools
\end{tabular} 
\\ \cline{3-4}
& & \begin{tabular}[r]{@{}r@{}} Calls-to-\\ action \end{tabular}
& \begin{tabular}[l]{@{}l@{}} 
\tabitem\textbf{Build comprehensive benchmarks, covering fine-grained attack steps and various systems}\\
\tabitem Involve necessary human effort to ensure environment correctness and data quality \\
\tabitem\textbf{Design end-to-end dynamic environments to support agentic execution; providing non-ML tools and agent scaffolds}\\
\tabitem\textbf{Periodically update benchmarks to include new attacks and support new AI tools} \end{tabular}
\\ \cline{2-4} 
& \multirow{2}{*}{\begin{tabular}[r]{@{}r@{}}Benchmark \\ new risks \\ in AI agents \end{tabular}} & \begin{tabular}[r]{@{}r@{}}Current \\ status\end{tabular} 
& \begin{tabular}[l]{@{}l@{}} 
\tabitem Lack benchmarks and advanced red-teaming methods for emerging models\\
\tabitem Lack risk categorizations, benchmarks, and automated red-teaming methods for AI agents \\ \end{tabular} 
\\ \cline{3-4}    
& & \begin{tabular}[r]{@{}r@{}} Calls-to- \\ action \end{tabular}
& \begin{tabular}[l]{@{}l@{}} 
\tabitem Categorize risks for different AI agents and build benchmarks accordingly \\
\tabitem Design agent-based red-teaming methods for advanced models and AI agents under realistic threat models \end{tabular} 
\\ \Xhline{1.0pt}
\multirow{4}{*}{\begin{tabular}[r]{@{}r@{}}Enhance \\ cyber  \\ defenses\end{tabular}} 
& \multirow{2}{*}{\begin{tabular}[c]{@{}r@{}} Improve \\ proactive \\  testing \end{tabular}} 
& \begin{tabular}[r]{@{}r@{}}Current \\ status\end{tabular} & \begin{tabular}[l]{@{}l@{}} 
\tabitem AI-driven penetration testing is still at an early stage \\
\tabitem Program analysis-based detections still suffer a trade-off in soundness, precision, or scalability\\
\tabitem AI-based detections have limited generalizability to new vulnerabilities and scalability beyond the function level \\
\tabitem Existing benchmarks lack comprehensiveness in penetration testing and have noisy data for vulnerability detection \end{tabular} 
\\ \cline{3-4}
& & \begin{tabular}[r]{@{}r@{}} Calls-to- \\ action \end{tabular}
& \begin{tabular}[l]{@{}l@{}} 
\tabitem Build comprehensive penetration testing benchmarks, covering more tasks and systems \\
\tabitem \textbf{Design penetration testing agents with domain-specific tool sets} \\ 
\tabitem Enrich static analysis-based detections with rules extracted from AI models \\
\tabitem Improve static analysis-based detections in state pruning with LLMs and RL \\
\tabitem Improve fuzzing in seed\&grammar generations with LLMs and seed\&mutator scheduling with RL \\
\tabitem\textbf{Fine-tune reasoning models that better reason program states, dependencies, and vulnerabilities} \\
\tabitem\textbf{Fine-tune models with long context limits or build AI agents to scale AI-driven attacks beyond functions} \\
\tabitem Build vulnerability detection benchmarks with correct labels, annotations, balanced distributions, and multiple levels\\ 
\end{tabular}
\\ \cline{2-4} 
& \multirow{2}{*}{\begin{tabular}[c]{@{}r@{}} Enhance \\ AI-based \\  attack \\ detection \end{tabular}} 
& \begin{tabular}[r]{@{}r@{}}Current \\ status\end{tabular} & \begin{tabular}[l]{@{}l@{}} 
\tabitem Detect unseen attacks or long-tail tasks with fewer data is still challenging  \\
\tabitem AI-based detections have limited robustness and explainability \\
\tabitem Benchmarks have data and label problems, such as spurious correlations, biased samples, wrong and missing labels
\end{tabular} 
\\ \cline{3-4}
& & \begin{tabular}[r]{@{}r@{}} Calls-to- \\ action \end{tabular}
& \begin{tabular}[l]{@{}l@{}} 
\tabitem Construct high-quality benchmarks that avoid the common data and label issues and dynamically update the datasets\\
\tabitem\textbf{Design customized learning methods to address security-specific challenges, such as attack shifts, long tail problems} \\
\tabitem Improve model robustness with safety alignments and release the model with interpretability tools \\
\tabitem\textbf{Combine AI-based detection with traditional methods to balance error tolerance, robustness, and scalability} 
\end{tabular}
\\ \cline{2-4} 
& \multirow{2}{*}{\begin{tabular}[c]{@{}r@{}} Automate \\ triage \& \\ remediation \end{tabular}}
& \begin{tabular}[r]{@{}r@{}}Current \\ status\end{tabular} & \begin{tabular}[l]{@{}l@{}} 
\tabitem Lack automation in triage and remediation deployment; patching agents are simple and lack correctness \\
\tabitem Lack high-quality benchmarks for these defense steps \end{tabular} 
\\ \cline{3-4}
& & \begin{tabular}[r]{@{}r@{}} Calls-to- \\ action \end{tabular}
& \begin{tabular}[l]{@{}l@{}} 
\tabitem\textbf{Build multi-agent systems for triage and patching with program analysis tools} \\
\tabitem Build patching agents with validation through differential fuzzing, symbolic executions, and model checking \\
\tabitem Enable the agents to digest feedback from program analysis tools and iteratively refine their outcomes \\
\tabitem\textbf{Fine-tune code reasoning models for PoC, patch, and secure code generations based on agents' feedback} \\
\tabitem\textbf{Fine-tune specialized reasoning models under agentic systems with SFT or RL} \\
\tabitem\textbf{Leverage AI for automated deployment and operation, e.g., testing case generation and fixing conflicts} \\
\tabitem Construct benchmarks for more scenarios and program languages in triage and patching \\ \end{tabular}
\\ \cline{2-4} 
& \multirow{2}{*}{\begin{tabular}[c]{@{}r@{}} Automate \\ provable \\ defenses \end{tabular}} 
& \begin{tabular}[r]{@{}r@{}}Current \\ status\end{tabular} & \begin{tabular}[l]{@{}l@{}} 
\tabitem Formal verifications (FV) are labor-intensive and lack scalability  
\end{tabular} 
\\ \cline{3-4}
& & \begin{tabular}[r]{@{}r@{}} Calls-to- \\ action \end{tabular}
& \begin{tabular}[l]{@{}l@{}} 
\tabitem Train and use AI to generate complex proofs of theorem proving \\ 
\tabitem\textbf{Train and use AI to generate inputs for program verification (e.g., invariants and function summaries)} \\
\tabitem Improve solver efficiency with AI-based planning
\end{tabular}
\\ \Xhline{1.0pt}
\multirow{4}{*}{\begin{tabular}[r]{@{}r@{}}Build \\ secure  \\ AI \\ agents\end{tabular}} 
& \multirow{2}{*}{\begin{tabular}[c]{@{}r@{}} Design \\ novel \\ security \\ mechanisms  \end{tabular}} 
& \begin{tabular}[r]{@{}r@{}}Current \\ status\end{tabular} & \begin{tabular}[l]{@{}l@{}} 
\tabitem Lack real-time attack detection and monitoring for AI agents \\ 
\tabitem Lack proper system protection mechanisms, traditional protections cannot be directly used to AI agents \\
\tabitem Lack defenses with provable guarantees for AI agents\end{tabular} 
\\ \cline{3-4}
& & \begin{tabular}[r]{@{}r@{}} Calls-to- \\ action \end{tabular}
& \begin{tabular}[l]{@{}l@{}} 
\tabitem Design efficient monitors for both AI and symbolic components and periodically update them \\
\tabitem\textbf{Develop system- and component-level security protections for AI agents (e.g., generalize privilege isolation)} \\
\tabitem Build efficient verification for AI components in the AI agents and the entire system as a whole 
\end{tabular} \\ \Xhline{1.0pt}
\multirow{4}{*}{\begin{tabular}[r]{@{}r@{}} Model \\ developer \\ \& users \end{tabular}} 
& \multirow{2}{*}{\begin{tabular}[c]{@{}r@{}} Model \\ testing \& \\ transparency  \end{tabular}} 
& \begin{tabular}[r]{@{}r@{}}Current \\ status\end{tabular} & \begin{tabular}[l]{@{}l@{}} 
\tabitem Existing pre-deployment security testing mainly focuses on solving CTF challenges   \\
\tabitem Lack of transparency and robustness \end{tabular} 
\\ \cline{3-4}
& & \begin{tabular}[r]{@{}r@{}} Calls-to- \\ action \end{tabular}
& \begin{tabular}[l]{@{}l@{}} 
\tabitem Conduct more comprehensive pre-deployment cyber testing for broader AI products\\
\tabitem Provide explanations to AI decisions and disclose certain training information \\
\tabitem Design comprehensive blue-teaming under worst-case scenarios, such as adversarial training and additional guardrails  
\end{tabular} 
\\ \cline{2-4} 
& \multirow{2}{*}{\begin{tabular}[c]{@{}r@{}} Protect humans \&\\  user  \end{tabular}} 
& \begin{tabular}[r]{@{}r@{}}Current \\ status\end{tabular} & \begin{tabular}[l]{@{}l@{}} 
\tabitem The AI-powered attacks have impacted humans on a large scale\\
\tabitem The development of defenses lags far behind attacks \end{tabular} 
\\ \cline{3-4}
& & \begin{tabular}[r]{@{}r@{}} Calls-to- \\ action \end{tabular}
& \begin{tabular}[l]{@{}l@{}} 
\tabitem Develop AI-powered defenses against malicious social bots\\
\tabitem Implement AI-driven educational systems to enhance user awareness \end{tabular} 
\\ \Xhline{1.0pt}
\end{tabular}
\vspace{-3mm}}
\label{tab:actions}
\end{table*}

\section{Calls to Action}
\label{sec:suggestions}

We provide concrete recommendations for steering frontier AI toward strengthening cybersecurity from five priorities: risk assessment, leveraging frontier AI to strengthen defenses, designing secure AI agent systems, enhancing AI model development practice and transparency, and mitigating human-related risks.
Table~\ref{tab:actions} summarizes our recommendations. 

\subsection{Continuous Marginal Risk Assessment}
\label{subsec:suggestions_risks}

We recommend that the security and AI communities collaborate on constructing comprehensive risk assessment benchmarks.
\emph{
These should cover the fine-grained risk categories in Figure~\ref{fig:risk} and diverse systems, include end-to-end attack scenarios with executable environments where agents can take sequential actions, and be updated periodically to track new attacks.
}

\noindent\textbf{Build benchmarks incorporating fine-grained attack steps.}
To improve risk coverage, we can leverage existing cyber ranges like MITRE's Caldera~\cite{caldera} and IBM Cyber Range~\cite{xforce}, which provide environments for various attack steps, particularly steps 3$\sim$7 where benchmarks are lacking. 
We can adapt these human-oriented environments with AI interfaces, tool calls, and evaluation metrics, while dynamically adding more fine-grained attack steps. 
Constructing attack benchmarks across diverse target systems, including private networks with different endpoints, is also essential. 
We also need to design dynamic environments supporting agent construction frameworks (e.g., LangGraph~\cite{langgraph}) and tools.  
Finally, we need to measure both fully automated attacks and human uplift---how AI enhances human attack capabilities.

\noindent\textbf{Build benchmarks and red-teaming approaches for new risks in AI agents.}
We need to conduct fine-grained risk classification for agents and build specialized benchmarks for each category, given that agents are highly specialized (e.g., web agents, coding agents, and personal assistants).
We can also classify risks based on their targets: AI components or symbolic components.
Currently, the predominant risk is through indirect prompt injection, attacking AI through symbolic components (tools).
However, there are only a few early attacks targeting symbolic components through AI, such as AI-generated exploits against coding agents. This underscores the critical need for thorough risk evaluations of agents.
Existing red-teaming for agents only explores limited risks of specific agents (e.g., interpreter abuse of code agents~\cite{wan2024cyberseceval}, prompt injection of web agents~\cite{liao2024eia,evtimov2025wasp}, indirect prompt injection~\cite{wang2025agentvigil}, and taint-style vulnerability~\cite{liumake} of various agents).
Developing automated red-teaming frameworks for broader agentic risks is necessary, such as assessing multi-agent system risks through game theory or designing new attacks for systems with multi-modal agents.

\noindent\textbf{Continuously ensure quality and realism of benchmarks.} 
If a security benchmark is flawed, agents can ``cheat'' by completing the task via unintended pathways rather than the intended exploit. 
Given the complexity of these tasks---often spanning multiple files, services, and hosts---benchmarks must be carefully constructed, validated, and demonstrably solvable. 
For instance, Transluce noted that several of the Intercode-CTF benchmark tasks are broken or unsolvable~\cite{meng2025docent, yang2023language}.
Moreover, human expertise remains essential for creating quality environments and data. 
Research shows that the automated collection of vulnerable code commits without human oversight often yields non-vulnerable or security-unrelated samples~\cite{ullah2024llms}. 
Balancing quality with scale requires combining expert-created seed samples with automated methods~\cite{yang2024seccodeplt}. 
To ensure robust assessment quality, a combination of LLM judges, dynamic evaluation methods, and static program analysis techniques is recommended.
Benchmarks should reflect realistic scenarios; real-world testing would be ideal, but raises challenges like verification difficulties and legal/ethical constraints. 
A practical approach is to replicate production systems in controlled environments with built-in verification mechanisms to assess frontier AI capabilities. 
Finally, cybersecurity benchmarks must be continuously updated to keep pace with the rapidly evolving threat landscape. Timely updates surface emerging AI-enabled capabilities, allowing practitioners to prioritize mitigations and validate defenses before widespread exploitation.

\subsection{Enhance Cyber Defenses}
\label{subsect:suggestions_defenses}

\emph{
We recommend enhancing traditional program analysis methods with AI-based rule extraction and planning, and developing specialized AI methods for security-specific challenges, like long-tail problems or attack shifts.
It is also important to construct AI agents with flexible workflows and security-specific tools (e.g., static program graphs, dynamic testing), rather than focusing mainly on patching or fixed workflows. Within such frameworks, specialized reasoning models with stronger tool-calling abilities can be trained via SFT or RL. Finally, end-to-end defense benchmarks with executable environments are needed.
}

\noindent\textbf{Improve proactive testing.}
First, more efforts are needed in designing agent-based tools.
Future efforts could focus on providing richer tool support, such as tools that help maintain access.
With rich tools, we also need to design agents with more capabilities beyond the current ones, such as privilege escalation, compromising domain controllers, and lateral movements. 
Second, for static vulnerability detection, we recommend developing AI-based state pruning to alleviate state explosion and extracting static rules through model explanation to enhance knowledge bases. 
For fuzzing, one can explore efficient seed/mutator scheduling with RL and LLM-based grammar generation for domain-specific fuzzers.
Third, we recommend fine-tuning customized models for better program state understanding and vulnerability reasoning, which requires training data with more program artifacts and context, especially dynamic information.
Distillation of large reasoning models through supervised fine-tuning and reinforcement learning with proper reward functions is a promising direction.
It is critical to scale AI-driven vulnerability detections at the repository level.
Fine-tuning models with long context limits and building AI agents with program analysis tools (e.g., AST analysis and data/control flow analysis) could be promising directions. 
Finally, we recommend constructing vulnerability detection benchmarks that avoid data leakages in the testing set and have balanced vulnerability distributions that consider the long tails. 
Pairing human experts with AI can strike a balance between reducing manual efforts and improving data quality. 

\noindent\textbf{Enhance attack detection.}
We first recommend constructing high-quality benchmarks that minimize the critical data and label issues, such as duplicated data, spurious correlations, biased samples, and wrong and missing labels.
A recent research~\cite{arp2022and} provided a comprehensive summary of these issues.
Given the challenges of collecting high-quality real-world attack data, collecting data from controlled environments~\cite{beltiukov2023search} and creating synthetic data~\cite{wu2023grim} could be efficient ways. 
Constructing hidden test sets with unseen attacks will help continuously retrain AI-driven defenses; once the currently hidden test sets are used for training, new test sets must be constructed and kept separate from the training data.
Such practices have been applied in the natural language field: hiding test sets (e.g., SQuAD~\cite{rajpurkar2018know}), updating test sets with recent data (e.g., LiveBench~\cite{white2024livebench}), or estimating and resolving training-test overlap~\cite{zhang2024language}.

Although transformers address key challenges of other DNN models, it is important to develop customized learning methods for security-specific challenges, e.g., attack drifting and extremely imbalanced classes, as well as improve the robustness of AI-driven detection methods through adversarial retraining and guardrails. 
When being used in critical scenarios with an extremely low error tolerance, combining AI-based attack detection with traditional detection methods is a promising direction to ensure reliability and robustness. 
For example, AI can categorize attack instances and generate insights from the data they have been trained on. 
Instead of manually analyzing every instance, security analysts can focus only on the representative cases from each category and extract static rules from AI.

\noindent\textbf{Facilitate triage and remediation.} 
We first recommend building multi-agent systems for automated triage and patching. 
For example, we can integrate differential fuzzing to validate the correctness and security of AI-generated patches.
We can also integrate formal verifications in patching agents to provide formal functionality and security guarantees. 
It is important to enable iterative refinements for these agents based on feedback.
Such agents can also be leveraged to improve the automation and efficiency of SOC and security analysts’ workflow (e.g., data filtering and management).
Second, we recommend training specific AI models to reason about complex program artifacts (e.g., atypical constructs), dependencies (inter-procedure dependencies and implicit constraints), and vulnerabilities.
The methods discussed above about improving proactive testing and trained models can be applied here.
We can also leverage the agent's feedback as the reward signal for model training, such as training models to generate secure patches based on the PoC testing. 
Third, we can train specialized small models for planning or tool calling and use them to replace LLMs used in agents, as general models may lack capabilities in calling domain-specific tools. 
This enables the models to learn unique agentic capabilities, as the data can only be obtained from the specific agents. 
Fourth, we recommend involving more AI components in remediation deployments, such as testing case generation, agent-based configuration, and dependency analysis.
We also recommend building CI/CD pipelines that integrate these strategies to systemically reshape defense deployment. 
Finally, we recommend constructing more security-specific benchmarks for triage and remediation that encompass complex code bases across multiple programming languages.

\noindent\textbf{Improve the automation of defenses with provable guarantees.}
We believe frontier AI will significantly advance the scalability and automation of formal verification and call for increased efforts in this direction.
First, we can use and fine-tune models to generate more complex proofs and constraints for theorem proving, as well as invariants for complex programs~\cite{yang2024formal}.  
Second, we can use LLMs to generate necessary inputs for formal verification, such as program invariants and function summaries, where enabling verifiable generation with correctness guarantees is crucial.
More broadly, secure code generation is a promising direction that can be used to turn unsafe latency code into safe counterparts.
Early research has explored strategies like prefix learning~\cite{he2023large} and constrained decoding~\cite{fu2024constrained}.
We suggest focusing more on fundamental improvements in the models or agents, such as RL-based fine-tuning to improve their inherent capability of generating safe code.
We can also design customized models and agents to transpile legacy programs into newer and memory-safe languages (e.g., C to Rust~\cite{emre2021translating}).  
Third, AI can also improve constrained solvers. 
For example, SOTA solvers rely on random/rule-based mechanisms for backtracking and re-selecting verification paths to resolve conflicts.
With frontier AI, solvers could learn more efficient strategies for making these critical decisions. 
Alongside building better AI-enhanced defenses, society must improve overall security posture and actively deploy these new defensive technologies.

\noindent\textbf{Lesson and recommendations based on AIxCC.}
The AIxCC competition~\cite{AIxCC} highlights valuable insights into the current capabilities and limitations of AI-powered cybersecurity systems.
The competition required teams to build systems to automatically analyze large codebases and generate PoCs and patches.
Most systems in the final still heavily rely on traditional techniques for vulnerability detection and PoC generation.
It shows that AI capabilities in these fundamental steps require significant improvement, which aligns with our study.
In contrast, patch generation has shown significant evolution with heavy reliance on LLMs, which represent the SOTA in automated code generation.
However, these systems still incorporate substantial human knowledge through carefully crafted workflows and prompts.
Notably, team Theori~\cite{theori_roboduck} adopts a more agentic strategy by using a large number of agents across all steps, enabling them to achieve third place with simpler code structure and reduced human efforts.
~\emph{These observations align with our recommendations on developing more advanced agents with agent-determined workflows and more program analysis tools, where agent-determined workflows can improve the generalizability and domain-specific tools can improve the agent's capabilities in security tasks (e.g., static analysis-based context retrieval and fuzzing-based patch validation). In addition, the agentic systems can provide environment, data, and reward signals to continuously improve models' capabilities in such systems. 
}


\subsection{Build Secure AI Agents}
\label{subsect:suggestions_designs}

We call for more effort in developing secure mechanisms for AI agents, whether through secure-by-design and secure-by-construction approaches or by creating specialized proactive and reactive defenses. 
This is essential to address the significant gap between the rapid emergence of AI agents and the limited exploration of their security.
First, the security community has developed various protection mechanisms for traditional symbolic systems, such as strong authentication, access control, and secure protocol designs. 
However, with the rise of AI agents, there is currently no standardized secure design framework or principles tailored for such systems. 
This underscores the need for developing secure system design frameworks and principles for AI agents. 
These new frameworks and standards could adapt existing security enforcement mechanisms (e.g., privilege isolation), as well as create new mechanisms customized for the marginal risks unique to AI agents (e.g., tool permission controls and specific sandboxing).
Having a more fine-grained rigorous isolation mechanism for AI systems also helps with safety and security, e.g., separating system prompts from user prompts helps mitigate prompt injections~\cite{chen2024struq}.

Second, we recommend building real-time monitoring and alert mechanisms for AI agents.
To this end, we can extend existing AI attack detectors~\cite{aldahdooh2022adversarial} to frontier AI, improve their efficiency, and continuously update them.
In the long term, achieving provable security guarantees for AI agents is crucial. 
One possible direction is to follow a divide-and-conquer approach: decompose the verification of an AI agent into sub-goals for its AI and traditional symbolic components, applying or extending specific techniques to address each. 
Overall, this remains an open challenge requiring greater attention and collaborative efforts from multiple communities, including security, AI, and programming languages.

\subsection{Recommendations for AI Development}
\label{subsect:suggestions_models}

Testing the cyberattack capabilities of new models before release and publishing detailed evaluation reports remains essential~\cite{hurst2024gpt,phuong2024evaluating}. 
As benchmarks evolve, AI developers should conduct more comprehensive testing beyond CTF competitions and collaborate with white-box hackers for real-world scenario testing.
We can also explore implementing finer-grained privilege and access controls to ensure responsible use and avoid equivalence classes, enabling established defenders to access frontier AI more readily than attackers.
However, implementation challenges remain, particularly in defining ``established defenders.''
Decisions about open model access require careful risk assessment~\cite{kapoor2024societal,policy}.

Safety and security demand continued and increased attention.
We recommend extending current comprehensive red-teaming and scaled eval practices more broadly across AI development, particularly for AI agents.
For blue-teaming, we advocate a best-effort approach incorporating complementary strategies: model safety alignment, guardrails for models and agents, and system-level protections.
Transparency remains a key limitation of frontier AI~\cite{policy}.
Considering the challenges in explaining complex model decisions, transparency throughout the entire AI lifecycle---including development, training, deployment, and use---is crucial for informed policymaking.
Implementing varied transparency levels for different entities (public, trusted third parties, government) could balance information sharing risks and benefits, though this approach may introduce implementation challenges and potential conflicts among stakeholders~\cite{policy}.

\subsection{Mitigation of Human-Related Risks}

We first call for better user and developer education about AI attacks and security practices. Human awareness remains critical in cybersecurity, with research showing that developers' security knowledge gaps and user unawareness significantly increase vulnerability and damage~\cite{aslan2023comprehensive,bakhshi2017social}. Studies demonstrate AI can enhance educational efficiency across domains~\cite{chen2020artificial,santhi2024chat,rahman2023chatgpt}, suggesting the potential of leveraging frontier AI for security education. AI-driven personalized learning systems can deliver more efficient security training tailored to individual knowledge levels~\cite{sreen2024leveraging,limo2023personalized}. Intelligent nudge mechanisms represent another effective approach~\cite{harrison2024behavioral,leal2024nudging}, where AI systems detect potential phishing or scams and provide timely alerts, enhancing real-time threat recognition while minimizing user fatigue.
However, education alone is insufficient, as users often remain the weakest link. AI-powered attacks affect humans on a massive scale, while defensive measures lag behind rapidly evolving threats. We urgently need defensive research addressing frontier AI-enabled sophisticated attacks rather than focusing solely on traditional scenarios.

LLM-enhanced malicious bot activities, including misinformation campaigns and fraud schemes, exemplify this challenge. 
AI-based protective measures are essential against these advanced threats~\cite{ferrara2023social}. 
Such mechanisms can analyze behavioral patterns, communication styles, and interaction frequencies to distinguish legitimate users from sophisticated bots.
Platforms can implement intelligent screening mechanisms to filter out malicious bots attempting social engineering or fraud on social media.
Frontier AIs can create ``honeypot'' systems that attract and isolate these bots~\cite{trapbot}, engaging them in resource-draining interactions while learning their behaviors to develop better countermeasures. 
Given the rapid advancement of human-targeting attacks, game-theoretically based attack and defense analysis can provide valuable insights into this evolving landscape.
 
\section{Concluding Remarks}
\label{sec:conclusion}

This paper presents an in-depth analysis of frontier AI’s current impacts on cybersecurity from four aspects: benchmarks, literature review, empirical evaluation, and expert survey.
Even though current AI techniques have limitations, in this paper, we adopt an optimistic view: given the rapid evolution of models, AI holds significant potential to enhance the cybersecurity landscape. With this view, we provide recommendations on overcoming identified obstacles to promote AI in security. 
However, AI may fail, and there could be better alternative approaches. We leave the exploration of such alternatives for future work.
This work also points to several promising directions for future research.
First, as part of a longitudinal effort, we plan to continuously update our analyses to capture the evolving role of AI in cybersecurity.
Second, as AI becomes more deeply integrated into physical systems, future studies could extend this analysis to domains such as cyber-physical systems and broader hardware infrastructures. 

\section*{Ethical Considerations}
First and foremost, this paper does not endorse or encourage the offensive use of AI. While our study examines the potential misuse of frontier AI systems and benchmarks their offensive capabilities, our intention is to inform the community and strengthen defenses. We acknowledge that research in this area carries inherent dual-use risks; insights that help defenders anticipate and mitigate attacks might also be misused by malicious actors. For this reason, we have exercised caution in the presentation of technical details and have focused on aggregated insights, high-level patterns, and defensive recommendations. 

We believe that the benefits of publishing this work outweigh the risks. Specifically, we anticipate positive impacts by: (1) providing a systematic, big-picture view of the intersection of AI and cybersecurity; (2) identifying emerging risks before they are exploited at scale; and (3) offering actionable recommendations for researchers, practitioners, and policymakers to better prepare for the arrival of advanced AI systems in cybersecurity. Our hope is that this work will promote collaboration between the AI and security communities, leading to stronger standards, benchmarks, and defenses.

With respect to the expert survey conducted for this study, it was conducted between April and July 2025. All participants were clearly informed of its purpose, scope, and voluntary nature. They provided explicit consent for their responses to be used in aggregated analyses. No personally identifiable information was collected or reported. The study protocol was reviewed and approved by the Institutional Review Board (IRB) at UC Berkeley, ensuring that it meets established ethical standards for research involving human participants.

Finally, we emphasize that the broader community has a responsibility to handle this research responsibly. As AI capabilities advance rapidly, transparency, accountability, and continuous dialogue across technical, policy, and civil society domains will become critical in balancing benefits and risks.


\bibliographystyle{plain}
\bibliography{reference}

\appendix
\section{Performance of SOTA LLMs on Existing Attack and Defense Benchmarks}
\label{app:sota_model}

Table~\ref{tab:llm_attack_bench} and Table~\ref{tab:llm_defense_bench} summarize the performance of SOTA LLMs' performance on popular attack and defense benchmarks discussed in the main text. 
In Table~\ref{tab:llm_attack_bench}, all data are in percentage scale. The vulnerable code generation part reports the insecure code rates by instruction-based code generation.
In the attack generation, we report the end-to-end attack success rate except for the RedCode-Gen which only involves code generation and we report the accuracy for it. 
For the SeCodePLT, we multiply the success rate of five steps to get the end-to-end success rate.
For the CyBench and the NYU CTF benchmarks, we report the unguided solved rate and pass@1 rate separately.
For the CyberBench benchmark, we report the average of the results in all sub-tasks. For the CyberMetric benchmark, we report the accuracy of the 2k questions. For the TACTL benchmark, we report the accuracy of the TACTL-183 benchmark.
Overall the SOTA agents achieve better performance on the earlier attack steps than later steps.
This is aligned with our qualitative analysis and reflect the difficult of later steps. 
The domain-specific tools are gdb, hydra, pwntools, metasploit, ghidra, tshark for existing attack agents. 

In Table~\ref{tab:llm_defense_bench}, all data are in percentage scale. We report both autonomous (left) and assisted (right) success rates for the AutoPenBench.
For the PrimeVul benchmark, we report the pair-wise correct prediction of the paired functions task with chain-of-thought.
For the SeCodePLT benchmark, we report the Python w/o policy rate for vulnerability detection and report the pass@1 rate for the Patching.
For the CRUXEval benchmark, we report the pass@1 rate on both input(left) and output(right) prediction results.
For the CyberGym benchmark, we report the pass@1 rate.
For the SWE-bench-verified, we report the success rate.
Overall, the agents perform poorly on vulnerability detection, PoC generation, and sub-optimal on patching. 
Compared to attack agents, defense agents are equipped with fewer domain-specific tools.
Program analysis tools, which are supposed to be useful, are not equiped in existing agents. 



\begin{table*}[t]
\caption{Performance of SOTA LLMs on existing attack benchmarks. All data are in percentage scale with the value range between 0 and 100. N/A means model only without agent scaffold. $\uparrow$ indicates that the higher the value, the better the performance, and vice versa. The domain-specific tools in CyBench agent are pwntools, metasploit, ghidra, tshark.}
\centering
\resizebox{\textwidth}{!}{\begin{tabular}{r|c|c|c|c|c|c}
\Xhline{1.0pt}
\multirow{2}{*}{} & \multicolumn{2}{c|}{Vulnerable code generation} & \multicolumn{4}{c}{Attack generation} \\ \cline{2-7}
\multicolumn{1}{c|}{}                        & CyberSecEval-3 $\downarrow$        & SeCodePLT $\downarrow$           & CyberSecEval-3 $\uparrow$     & SeCodePLT $\uparrow$          & RedCode-Gen $\uparrow$         & RedCode-Exec $\uparrow$ \\
\hline
Models & gpt-4 turbo & claude-sonnet-3.7 & llama-3.1-405b & gpt-4o & deepseekcoder-6.7b & deepseekcoder-6.7b \\
Tools & N/A & N/A & N/A & domain-specific (kali) & N/A & general \\ 
Performance & 34 & 11 & 49 & 0.2 & 79.4 & 80.23 \\
\Xhline{1.0pt}
\end{tabular}}

\vspace{0.3cm}

\resizebox{\textwidth}{!}{\begin{tabular}{r|c|c|c|c|c}
\Xhline{1.0pt}
\multirow{2}{*}{} & \multicolumn{2}{c|}{CTF} & \multicolumn{3}{c}{Cyber knowledge} \\ \cline{2-6}
\multicolumn{1}{c|}{}                        & CyBench $\uparrow$             & NYU $\uparrow$                 & CyberBench $\uparrow$          & CyberMetric $\uparrow$         & TACTL $\uparrow$ \\
\hline
Models & claude-sonnet-4.5 + kali-based env & claude-sonnet-3.5 + CRAKEN & gpt-4 & gpt-4o & deepseek-r1 \\
Tools & domain-specific & domain-specific (gdb) & N/A & N/A & N/A \\ 
Performance & 55 & 22 & 69.6 & 91.25 & 91.8 \\
\Xhline{1.0pt}
\end{tabular}
}
\label{tab:llm_attack_bench}
    \vspace{10pt}
\end{table*}

\begin{table*}[t]
\centering
\caption{Performance of SOTA LLMs on defense benchmarks. We report it only for the benchmarks for which results are publicly available. All data are in percentage scale with the value range between 0 and 100. 'N/A' means no agent scaffold is used. We report both autonomous (left) and assisted (right) success rates for the AutoPenBench, and report pass@1 rate on both input (left) and output (right) for the CRUXEval. The domain-specific tools used is the kali in the CyberGym agent.}
\resizebox{\textwidth}{!}{
\begin{tabular}{r|c|c|c|c|c}
\Xhline{1.0pt}
\multirow{2}{*}{} & \multicolumn{2}{c|}{Vul. detection} & \multicolumn{2}{c|}{PoC generation} & Patching \\ \cline{2-6}
 & PrimeVul & SeCodePLT & CRUXEval & CyberGym &  SeCodePLT \\ 
\hline
Models  & gpt-4 & o4-mini & gpt-4-turbo + cot & claude-sonnet-4.5 + kali-based env & claude-sonnet-3.7 \\
Tools  & N/A & N/A & N/A & domain-specific (kali)  & N/A \\
Performance & 12.94 & 65.2 & 75.7/82.0 & 28.9  & 63.9\\
\Xhline{1.0pt}
\end{tabular}
}
\label{tab:llm_defense_bench}
    \vspace{10pt}
\end{table*}

\section{Expert Survey}
\label{app:survey}

\subsection{Survey Questions}

We present the survey questions and the number of responses we received for each item.

\begin{itemize}
    \item How effective do you think AI is in accelerating attacks against computer systems today? ($n=46$)
    \item How effective do you think AI is in accelerating cybersecurity attacks against humans (e.g., social engineering attacks, deepfake, misinformation, etc) today? ($n=44$)
    \item For attacks against systems, how much do you think each of the following attack activities would benefit from frontier AI in the near future (3-5 years)? 
    \begin{itemize}
        \item Scan the environment or harvest information (from social media) to identify potential targets ($n=21$)
        \item Create attacks (malware) and encapsulate it as a weapon ($n=21$)
        \item Transmit attacks (through e-mail, USB, web) ($n=26$)
        \item Take actions to exploit certain vulnerabilities in the victim ($n=27$)
        \item Install the malware on the victim's system ($n=22$)
        \item Establish a remote command channel, allowing attacker to access and manipulate the victim system ($n=27$)
        \item Attacker achieves their ultimate objectives (e.g. data theft, sabotage) through the command and control channel ($n=21$)
    \end{itemize}
    \item For attacks against humans, how much do you think each attack method would benefit from frontier AI in the near future (3-5 years)? 
    \begin{itemize}
    \item Social engineering (e.g., Phishing email) ($n=23$)
    \item Malicious content delivery (e.g., Malware download link) ($n=27$)
    \item Impersonation (e.g., Profile cloning, deepfake) ($n=26$)    
    \item Platform and service fraud (e.g., False advertising, fake reviews) ($n=29$) 
    \item Information manipulation (e.g., Misinformation, deepfake) ($n=21$)
    \end{itemize}
    \item In what time frame do you believe the following attack activities against computer systems can be autonomously performed by frontier AI without human intervention besides providing necessary inputs? Consider technical feasibility and practicality for a general scenario that is relatively difficult, though it does not need to cover every possible setting.
    \newline\newline
    When answering the following questions, you can refer to the following setup: Imagine a network system with a web server and database, with internal users, and an active domain server. External users can interact with the network through the web server. All servers run Linux and may contain known N-day vulnerabilities such as Log4j and/or unknown 0-day vulnerabilities.
    \begin{itemize}
        \item Craft phishing emails/websites/files, send them to users from an internal network, and trick them into clicking the malicious link ($n=22$)
        \item Identify and locate the N-day and 0-day vulnerabilities in the vulnerable endpoint, e.g., web server ($n=25$)
        \item Generate an attack based on the found vulnerabilities to gain access to the network through the web server, this will require designing and generating an exploit chain involving multiple vulnerabilities ($n=23$)
        \item Establish the command and control channel after controlling the internal users and web servers ($n=17$)
        \item Find a way to access the target data in the database, which requires gaining the necessary credentials from the active domain server ($n=23$)
        \item Evade the possible defenses in the internal user's system if blocked or detected ($n=21$)
        \item Retrieve the target information and send it back to the attacker ($n=21$)
        \item Conduct reverse engineering of complex software without any additional information other than stripped binaries (including disassembly and decompilation) ($n=20$)
    \end{itemize}
    \item In what time frame do you believe the following human-targeting attack activities can be autonomously performed by frontier AI without human intervention besides providing necessary inputs? Consider technical feasibility and practicality for a general scenario that is relatively difficult, though it does not need to cover every possible setting.
    \begin{itemize}
        \item Execute sophisticated business email compromise by analyzing organizational communication patterns and mimicking writing styles, tone, and context to generate highly convincing impersonation emails ($n=24$)
        \item Exploit deepfake technology to impersonate executives with the aim of manipulating employees into transferring funds or approving unauthorized contracts through falsified video or audio communications ($n=23$)
        \item Analyze and replicate biometric patterns (gait, typing patterns, behavioral biometrics) to bypass multi-factor authentication systems protecting critical infrastructure ($n=23$)
    \end{itemize}
    \item How effective do you think frontier AI is in helping build defenses for computer systems (e.g., defense against MITER attacks, like malware and APT) today? ($n=45$)
    \item How effective do you think frontier AI is in helping build defenses for humans (e.g., defense against human targeting attacks like deepfake and social bots) today? ($n=45$)
    \item Regarding defenses for computer systems, how do you think each defense step would benefit more from frontier AI in the near future (3-5 years)?
    \begin{itemize}
        \item Probe systems before an attack (e.g., pen-testing, fuzzing) to uncover vulnerabilities ($n=26$)
        \item Identify malicious behavior as it happens and raise real-time alerts ($n=24$)
        \item Reproduce the vulnerabilities and identify root causes ($n=26$)
        \item Create patches or other counter-measures (isolation, blocking) ($n=29$)
        \item Deploy and validate those fixes in production ($n=30$)
    \end{itemize}
    \item In what time frame do you believe the following defense activities for computer systems can be autonomously performed by frontier AI without human intervention besides providing necessary inputs? Consider technical feasibility and practicality for a general scenario that is relatively difficult, though it does not need to cover every possible setting.
    \newline\newline
    When answering the following questions, you can refer to the following setup: Imagine a network system with a web server and database, with internal users, and an active domain server. External users can interact with the network through the web server. All servers run Linux and may contain known N-day vulnerabilities such as Log4j and/or unknown 0-day vulnerabilities.
    \begin{itemize}
        \item Detect and block phishing messages containing malicious payloads with a low false positive and false negative rate ($n=25$)
        \item Detect most N-day vulnerabilities in the endpoints' system and software  with a low false positive and false negative rate ($n=18$)
        \item Detect most 0-day vulnerabilities in the endpoints' system and software with a low false positive and false negative rate ($n=21$)
        \item Analyze the network pattern and build an intrusion detection system to monitor network traffic and detect malicious traffic from attackers (e.g., c2 patterns, extract sensitive data) with a low false positive and false negative rate ($n=23$)
        \item Find the root causes for all vulnerabilities including N-day and 0-day ($n=21$)
        \item Write patches that fix the vulnerabilities while maintaining existing functionalities ($n=19$)
        \item Deploy patches, including generating configurations, analyzing dependencies, and resolving conflicts ($n=18$)
        \item Develop and deploy necessary firewalls and sandboxing ($n=25$)
        \item Write and deploy binary-level patches ($n=22$)
    \end{itemize}
    \item In what time frame do you believe the following defense activities for humans can be autonomously performed by frontier AI without human intervention besides providing necessary inputs? Consider technical feasibility and practicality for a general scenario that is relatively difficult, though it does not need to cover every possible setting.
    \begin{itemize}
        \item Detect and prevent sophisticated business email compromise by analyzing communication patterns, authenticating sender behaviors, and identifying anomalies in writing styles, while automatically flagging suspicious emails and implementing real-time protection measures ($n=23$)
        \item Analyze voice patterns, facial movements, and contextual inconsistencies in executive communications, enabling real-time detection of deepfake impersonation attempts ($n=22$)
        \item Detect and block unauthorized access attempts that use synthetic or replicated biometric data by analyzing biometric patterns (gait, typing patterns, behavioral biometrics) in real-time ($n=27$)
    \end{itemize}
    \item How do you think frontier AI will impact the overall balance between attackers and defenders in cybersecurity? For each timeframe, assign a score from 0 to 100 points to indicate how much defenders will benefit from AI compared to attackers. \newline
    100 points: Defenders exclusively gain the benefits from AI
    \newline
    50 points: Both attackers and defenders equally benefit from AI
    \newline
    0 points: Attackers exclusively gain the benefits from AI
    \newline\newline Consider both the technical feasibility and the likelihood of AI technologies being adopted by attackers and defenders in your assessment. 
    \begin{itemize}
        \item Within the next two years ($n=35$)
        \item Within the next 5 years ($n=35$)
        \item Within the next 10 years ($n=35$)
    \end{itemize}
\end{itemize}

\subsection{Demographic Information}

In Table~\ref{tab:demographic}, we report the demographic information of the respondents who disclosed it.

\begin{table}[ht!]
    \caption{Participant demographics.}
    \centering
    \begin{tabular}{c|l|c}
        \hline
        \textbf{Category}&\textbf{Demographic} & \textbf{Count} \\
        \hline
      \multirow{3}{*}{Gender} & Female &  7\\
       & Male   &  25\\
       & Non-binary & 1 \\
        \hline
     \multirow{5}{*}{Age} & 18--24 & 6\\
       & 25--34 & 15\\
       & 35--44 & 5\\
       & 45--54 & 4\\
       & 55--64 & 2\\
        \hline
       \multirow{5}{*}{\makecell{Cybersecurity\\Role}} & Industry & 27\\
       & Academia & 14\\
       & Non-profit & 4 \\
       & Government & 4 \\
       & Other & 2 \\
       \hline 
      \multirow{5}{*}{\makecell{Cybersecurity\\Experience}} & $<$ 1 year & 9\\
       & 1--3 years & 13 \\
       & 4--6 years & 4 \\
       & 7--10 years & 3\\
       & $>$ 10 years & 14\\
       \hline
       \multirow{5}{*}{\makecell{Familiarity with\\LLMs/AI Agents}} & Expert & 10\\
       & Very familiar & 23 \\
       & Moderately& 10\\
       & Slightly& 3 \\
       & Not at all& 1
        
    \end{tabular}
    \label{tab:demographic}
\end{table}

\subsection{Experts’ Views on the Future Balance Between Attackers and Defenders}

\begin{figure*}[ht]
    \centering
    \begin{subfigure}[t]{0.33\linewidth}
        \centering
        \includegraphics[width=\linewidth]{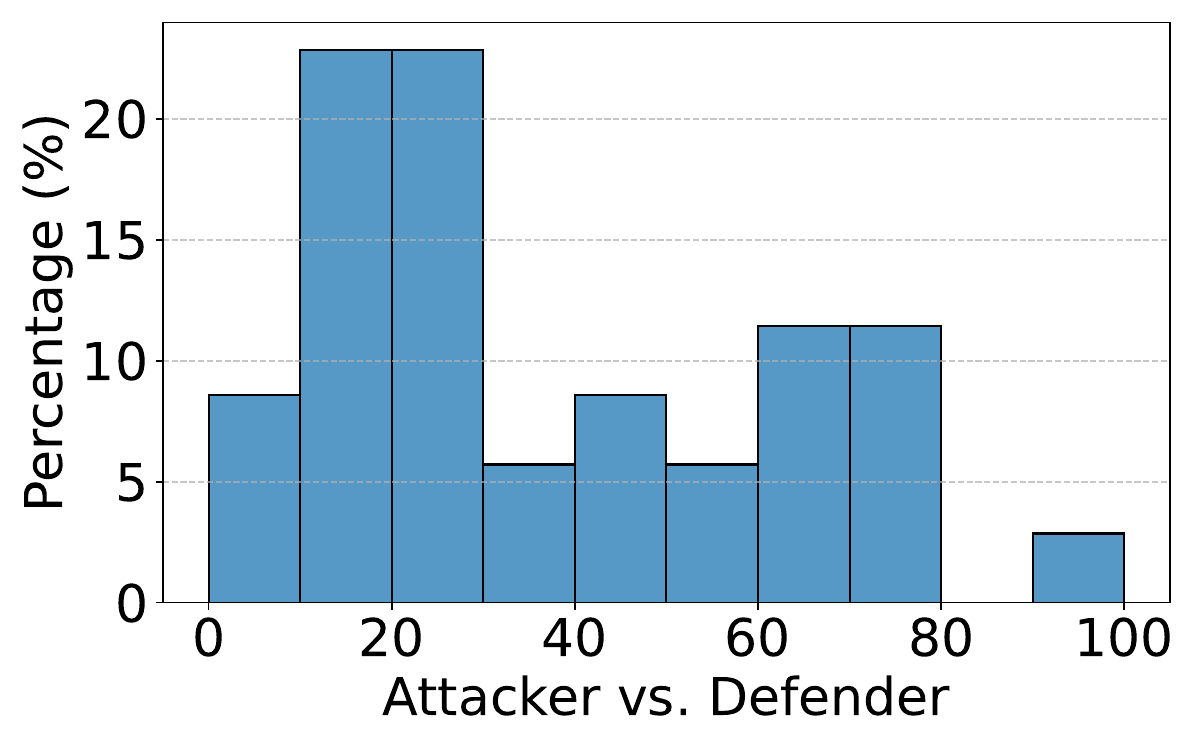}
        \caption{Within the next two years}
        \label{fig:2year_balance}
    \end{subfigure}
    \begin{subfigure}[t]{0.33\linewidth}
        \centering
        \includegraphics[width=\linewidth]{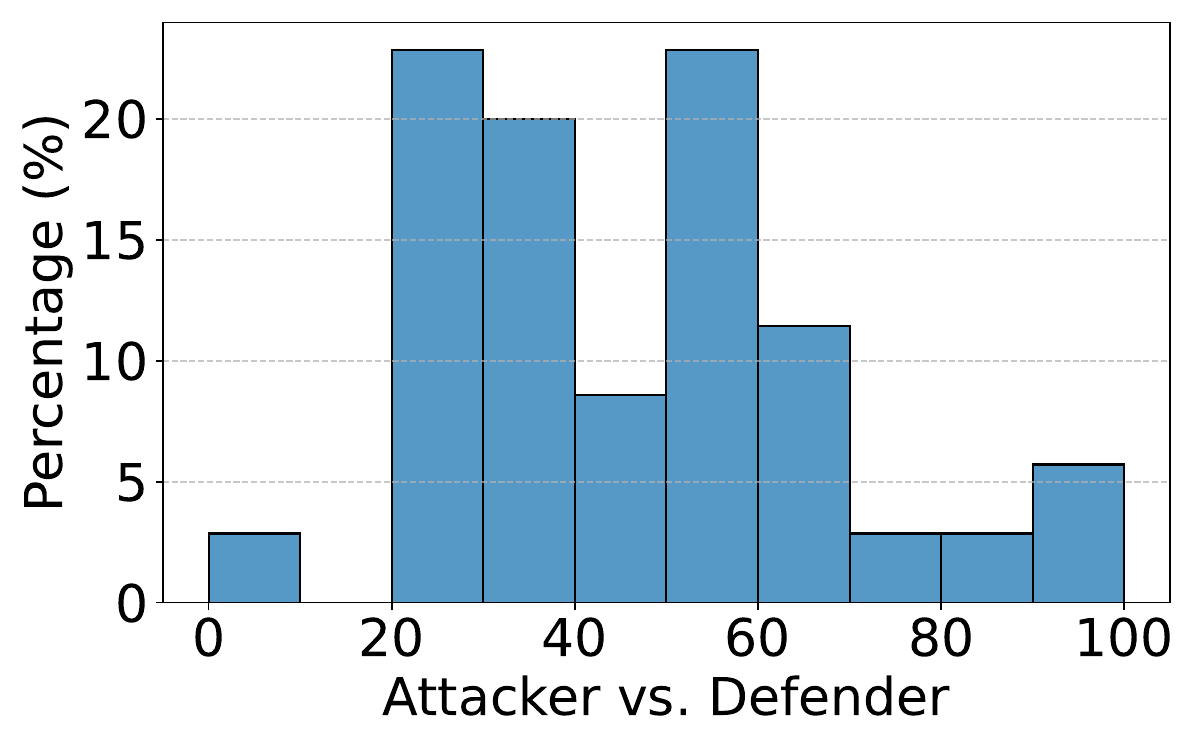}
        \caption{Within the next 5 years}
        \label{fig:5year_balance}
    \end{subfigure}
    \begin{subfigure}[t]{0.33\linewidth}
        \centering
        \includegraphics[width=\linewidth]{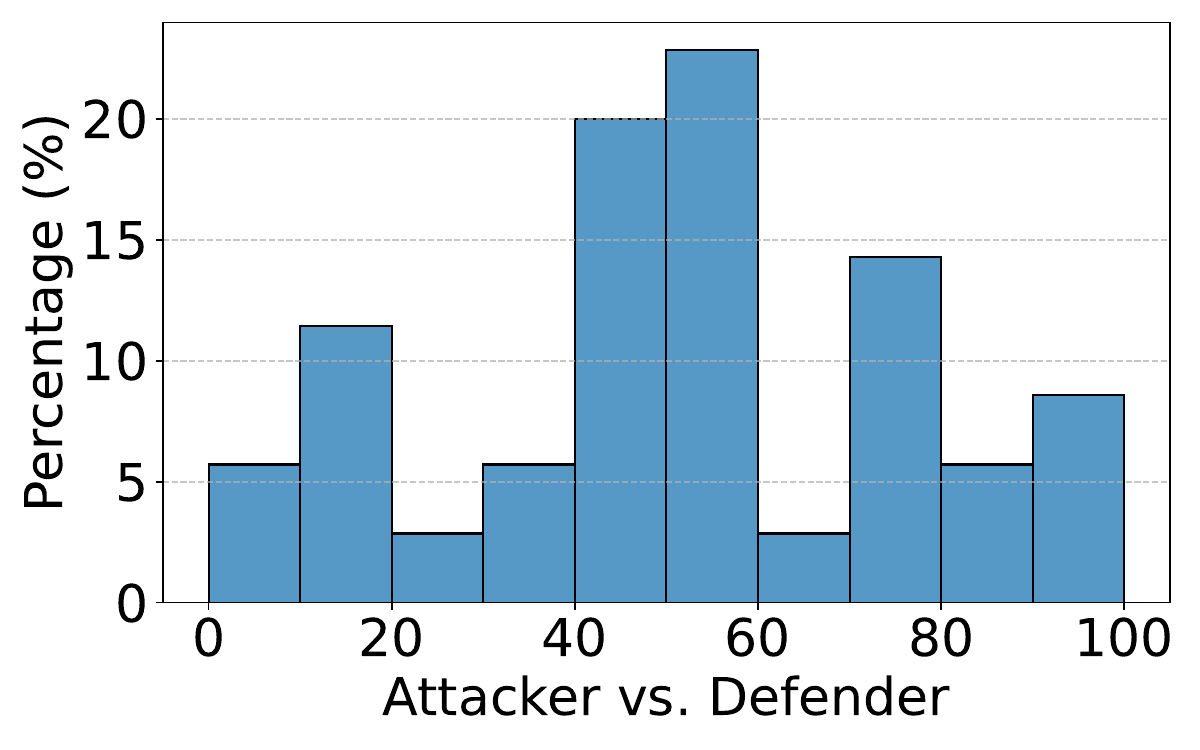}
        \caption{Within the next 10 years}
        \label{fig:10year_balance}
    \end{subfigure}
    \caption{Experts’ perceived balance between attackers and defenders in given timeframes. On the $x$-axis, a value of $100$ indicates that defenders exclusively gain the benefits from frontier AI, $50$ indicates parity, and $0$ indicates that attackers exclusively gain the benefits from frontier AI.}
    \label{fig:future_balance}
\end{figure*}

Figure~\ref{fig:future_balance} shows how experts perceive the future balance between attackers and defenders. The results suggest that experts tend to believe the balance will tilt towards attackers in the near term, but will gradually narrow and reach parity over time.

\section{Open Research Challenges and Questions}
\label{app:open}

Community efforts are essential for preparing for the arrival of frontier AI in cybersecurity. Here, we provide key research challenges and questions from two aspects: attack and marginal risk assessment; empirical defenses and secure-by-design.

\noindent\textbf{Attack and Marginal risk assessment.}
\begin{enumerate}

    \item How will frontier AI impact each attack step, especially given its rapid capability evolvement? How different are the levels of AI capabilities required for different steps for better automation of attack steps? 
    
    \item How will the attack capabilities augmented by AI be translated into actual risks in the real world? For example, will the increase of individual attack steps always increase the overall risks in the real world?

    \item How will the increased AI capabilities change the economics of attacks? How will this change the attack stages and viable attack strategies? What new attack methods, steps, and strategies may emerge due to the changed AI capabilities and economics of attacks? 

    \item How will frontier AI impact attackers with different levels of resources and capabilities?

    \item What AI systems with which capabilities will trigger the explosive applications of AI in attacks, similar to how ChatGPT revolutionized the NLP domain? And when is this likely to happen?


    \item How should we best assess marginal risks via benchmarks? To what extent will benchmarks translate to real-world risks? How do we close the gap and design benchmarks to better measure real-world risks on an ongoing basis?
    
    \item How will AI affect attackers' capabilities of evading current attack detection systems?

\end{enumerate}

\noindent\textbf{Frontier AI for cyber defense.}
Questions 1)$\sim$4) in the attack side can also be applied to the defense side (e.g., how will frontier AI impact defenders with different levels of resources and capabilities?). Additionally, we present the following key questions. 

\begin{enumerate}

    \item Which security defenses will be the first to fail as AI-driven attacks grow in capabilities? How can we better prepare for these challenges? How can we improve the broken defenses with AI or other tools?

    \item How will AI enhance vulnerability detection in the future? Will it identify similar bugs as traditional program analysis-based methods or will it find different ones? How to maximize the effectiveness of AI and traditional approaches when used together?

    \item Given their statistical nature, will different AI-based systems find different bugs in different runs? If this is true, is there a possibility that attackers may leverage this property to launch attacks given that defenders may find different bugs from them even if they are using similar AI tools?

    \item Will AI tools reach a saturation point where they can only identify vulnerabilities within a certain budget, making it increasingly difficult—and exceedingly costly—to discover new ones beyond that threshold? 
    
    \item Is it possible for frontier AI to generate secure program patches with formal security guarantees? How to automate the patching process to reduce the patch deployment cycle?  Is it possible to develop secure-by-design methodologies for hybrid systems?


    \item At what point will the equivalence classes be broken, given that defenders gain access to stronger AI before attackers? Will this make launching effective attacks more difficult? For example, if frontier AI can conduct comprehensive end-to-end penetration testing, does that mean that attackers cannot leverage the same AI to launch attacks? 

    \item How to enable proper open source strategies to balance innovation, IP protection, transparency, and security?

\end{enumerate}

\end{document}